\documentclass[
reprint,
superscriptaddress,
amsmath,amssymb,aps,pra,
]{revtex4-2}
\usepackage{xcolor}
\usepackage{xfrac}
\usepackage{graphicx}
\usepackage{dcolumn}
\usepackage{bm}
\usepackage{amsmath}
\begin{document}

\preprint{APS/123-QED}

\title{Effective Rabi frequency in semiconductor lasers and the origin of self-starting harmonic frequency combs}%

\author{Carlo Silvestri}
\affiliation{Institute of Photonics and Optical Science (IPOS), School of Physics, The University of Sydney, NSW 2006, Australia
}
\author{Franco Prati}%
\affiliation{Dipartimento di Scienza e Alta Tecnologia,
Universit\`a dell’Insubria, 22100 Como, Italy
}
\author{Massimo Brambilla}
\affiliation{ Dipartimento Interateneo di Fisica, Politecnico di Bari and CNR-IFN, UOS Bari, Italy
}
\author{Mariangela Gioannini}
\affiliation{Dipartimento di Elettronica e Telecomunicazioni, Politecnico di Torino, 10129 Torino, Italy
}
\author{Lorenzo Luigi Columbo}
\affiliation{Dipartimento di Elettronica e Telecomunicazioni, Politecnico di Torino, 10129 Torino, Italy
}

\begin{abstract}
Optical frequency combs have become a key research topic in optics and photonics. A peculiar comb state is the harmonic frequency comb (HFC), where optical lines are spaced by integer multiples of the cavity's free-spectral range. The spontaneous formation of HFCs has recently been observed in semiconductor lasers with fast gain recovery, such as Quantum Cascade Lasers (QCLs), although the underlying physical mechanism remains unclear. In this work, we provide a physical interpretation for the formation of HFCs in QCLs, based on a resonance phenomenon between an effective Rabi frequency and a mode of the laser cavity. 
This is corroborated by the results of the numerical integration of the effective semiconductor Maxwell-Bloch equations used to describe the multimode laser dynamics, as well as by the linear stability analysis of the continuous wave emission at threshold.

\end{abstract}
\maketitle
Optical frequency combs (OFCs) are dynamical regimes consisting of equally spaced optical lines, which adhere to a fixed phase relation \cite{Hansch_Nobel}. In the time domain, they correspond to a periodic signal generally in the form of optical pulses.
OFCs serve as a precise frequency ruler, making them pivotal for high-precision measurements. In fact, their introduction has driven significant advancements in the fields of metrology, spectroscopy, and optical communications \cite{spectrev, spect2012, Fortier2019, spect2017,perego2024optical}.

The last decades have witnessed a revolution in the study and development of comb sources, marked by the emergence of two typologies of devices with significant impact on chip-scale integration: optically pumped Kerr microresonators \cite{kippenbergrev} and fast semiconductor lasers, such as quantum cascade lasers (QCLs) \cite{Faist_1994} and quantum dot lasers (QDLs) \cite{Lu09}. These systems share the same underlying mechanism for comb generation, based on an efficient cascading four-wave mixing process \cite{Hugi2012,Faist_2016,Kippenberg2018,Bardella17}.

QCLs are unipolar semiconductor lasers emitting in the mid-infrared (mid-IR) and terahertz (THz) region of the electromagnetic spectrum, characterized by a picosecond carrier lifetime dominated by fast phonon scattering that makes QCLs ultrafast devices \cite{Faist_book}. This property, together with a small but finite value of the linewidth enhancement factor $\alpha$ (LEF)  which characterizes semiconductor active media, allowed for the demonstration of frequency combs in QCLs in the mid-IR and in the THz regions \cite{Hugi2012,Burghoff2014,Faist_2016}. This evidence has not only paved the way for transferring the aforementioned application advances to these spectral regions \cite{Villares_2014,SilvestriReview,Faist_2016}, but has also held significant importance from a fundamental perspective. In fact, QCLs exhibit a variety of comb states whose distinctive properties have been progressively clarified through theoretical advancements in recent years. In particular, QCL combs in the Fabry-Perot (FP) configuration display a coexistence of shallow amplitude modulation (AM) and frequency modulation (FM) behavior \cite{Optica_Faist_18, Cappelli2019, Silvestri20,Opacak2019, Burghoff20, Piccardo_PRL}. 
On the other hand, in the unidirectional ring configuration,
OFCs are associated with localized and global structures in the form of solitons and Turing rolls \cite{Columbo2018,NaturePiccardo,Bomeng1,Bomeng2,Columbo2021, NB2024} in analogy with the case of Kerr microresonators \cite{Pasquazi2018,Kippenberg2018}.
The formation of OFCs close to the lasing threshold is here accurately described by the prototypical Complex Ginzburg-Landau Equation (CGLE) \cite{NaturePiccardo,NB2024,Columbo2021}.

Another peculiar feature of QCLs is their ability to emit self-starting harmonic frequency combs (HFCs), characterized by an optical line spacing that is an integer multiple of the cavity FSR. The first observation of harmonic states in QCLs dates back to 2016 in FP mid-IR devices \cite{Mansuripur2016}, and the coherent nature of these regimes was demonstrated the following year \cite{Kazakov2017}. Subsequently, HFCs have also been reported in the THz range \cite{Wang2020,ForrerHFC} and in the ring configuration \cite{Jaidl21}, emerging as characteristic states of QCLs with compelling applications in microwave generation, pump-probe spectroscopy, and broadband spectroscopy \cite{PiccardoHFCOptex}. This has motivated the development of various configurations to induce and control HFCs with external elements, such as radiofrequency \cite{Dhillon1} and optical injection \cite{PiccardoOptical}, optical feedback \cite{Silvestri23,SilvestriPRA2025}, tuning of the laser temperature \cite{PiccardoHFCOptex}, surface and intracavity defects \cite{Riccardi24, Kazakov2021}.

Although many aspects of OFC formation in QCLs have been understood from a theoretical perspective, the physical origin of self-starting HFCs remains not completely understood \cite{SilvestriReview,PiccardoReview}. 
Early theoretical studies on the multi-mode instability underlying frequency comb formation in a FP laser were based on Maxwell-Bloch equations for two-level lasers.  Spatial hole burning (SHB), linked to the presence of a standing wave pattern and the associated carrier grating in the laser cavity, was considered instrumental for generating multimode emission close to the lasing threshold \cite{gordon,Boiko1, Boiko2}. It was shown that in a FP laser the single-mode state can become unstable in two ways: by a phase (FM) instability or by an amplitude (AM) instability and in this framework, HFCs in QCLs, were associated with the existence of peaks in the nonlinear parametric gain displaced by multiples of the cavity FSR \cite{Mansuripur2016,Belyanin2020}. 
This has some analogies with the well known Risken-Nummedal-Graham-Haken (RNGH) instability affecting two-level lasers and consisting in the amplification of the system Rabi frequency when it is resonant with a cavity mode \cite{Risken, Nos}. With our notations, the Rabi frequency is proportional to $\sqrt{X}$, where $X$ is the dimensionless field intensity in the stationary state. 
All these studies were based on the assumption of laser operating close to the threshold, which allows for keeping only the first harmonics in the population grating.
In \cite{Bassi2021,Lugiato2023} that assumption was removed, and this allowed to clarify that the RNGH instability in a FP laser occurs only for extremely high and unrealistic values of the pump and the instability close to threshold has a different origin. Indeed, the boundary of the instability domain observed close to threshold scales as $X^{1/4}$ instead of $\sqrt{X}$ \cite{gordon,Lugiato2019}.

In this letter, we study the phenomenon of HFC formation using a set of Effective Semiconductor Maxwell Bloch Equations (ESMBEs) \cite{Prati2007,Columbo2018} for a unidirectional ring QCL where
both the absence of SHB and periodic boundary conditions simplify system dynamics and allow for an analytical approach. The QCL ring configuration is in fact of particular interest for integrated photonic applications \cite{NB2024}.

ESMBEs realistically accounts for the gain structure of a semiconductor laser, characterized by a non-zero $\alpha$ factor which provides the well-known asymmetries in the medium gain and dispersion. We remark again that these mechanisms were key to the prediction and explanation of OFC formation close to threshold in unidirectional ring QCLs \cite{Columbo2018}. Moreover, in \cite{Columbo2018} homoclons typical of the Benjamin-Feir (BF) instability \cite{Aranson} were reported for $\alpha>1$ and close to threshold. Here we confirm and extend these results, but we also show that for $\alpha < 1$, there is an RNGH-type instability further from the threshold associated with an undamped effective Rabi frequency.

On this basis, we provide a theoretical explanation for the origin of self-starting HFCs. In particular we show that HFC spontaneously emerge from a resonance between a cavity mode and an intrinsic oscillation frequency of the material variables (carrier density and macroscopic medium polarization). This frequency shows a dependence on the square root of the intensity, a reknown fingerprint of the Rabi frequency in several coherent matter-field phenomena, so that we named it effective Rabi frequency (ERF). Remarkably, we could show that the resonance triggers a continuous-wave (CW) instability, leading to the spontaneous onset of an HFC whose order is equal to the number of cavity modes (or FSR) between the ERF and the reference frequency.
Our theoretical explanation is corroborated by a linear stability analysis (LSA) of the full ESMBE model that shows good agreement between the ERF and the frequency of maximum parametric gain. Our numerical simulations confirm and substantiate the predictions of the previous analyses. 

Similar results on HFC formation were predicted recently in unidirectional ring QDLs at telecom wavelengths using a two level system approach and inhomogeneous broadening \cite{Columbo18Qdot}.

We start from the Effective Semiconductor Maxwell-Bloch Equations that describe the dynamics of a multimode unidirectional ring QCL \cite{Columbo2018}:
\begin{align}
\frac{\partial F}{\partial \eta} + \frac{\partial F}{\partial t} &= \sigma[-F - P], \label{Eeq} \\
\frac{\partial P}{\partial t} &= \Gamma(1 + i\alpha)[-P - (1 + i\alpha)DF], \label{Peq} \\
\frac{\partial D}{\partial t} &= b[\mu - D + \frac{1}{2}\left(F^* P + F P^*\right)]. \label{Deq}
\end{align}
Here, the dynamical variables $F$, $P$, and $D$ represent the scaled electric field envelope, polarization, and carrier density, respectively. The field and carrier dynamical rates, normalized with respect to the polarization dephasing time $\tau_d$, are respectively $\sigma=\sfrac{\tau_d}{\tau_p}$ (where $\tau_p$ is the photon lifetime) and $b=\sfrac{\tau_d}{\tau_e}$ (where $\tau_e$ is the carrier lifetime), while $\Gamma$ corresponds to the scaled gain bandwidth. 
The parameter $\alpha$ denotes the linewidth enhancement factor. 
Finally, $\eta$ and $t$ are the dimensionless spatial and temporal variables, respectively. The scaling and the sign conventions follow the notation reported in \cite{silvestri2024unified}. Using this scaling, we convert the output power, time, and frequency into physical units, as presented in the figures of this work. The dynamical equations are complemented by the boundary condition of the ring cavity:
\begin{eqnarray}
F(0,t)&=&\sqrt{R}F(L,t),\label{bcring}
\end{eqnarray}
where $L$ is the cavity length and $R$ is the power reflection coefficient. In this work we assume the limit value $R=1$, i.e. a configuration similar to that reported in \cite{NaturePiccardo}.\\Differently from the Maxwell-Bloch equations for two-level systems, Eqs.~\eqref{Eeq}-\eqref{Deq} include essential aspects of the light-matter interaction characteristic of semiconductor lasers, such as the asymmetric frequency-dependent gain and refractive index and the coupling between field amplitude and phase, described by a non-zero $\alpha$ factor. These elements prove again a fundamental mechanism for the emergence of locking and, in this instance, of the HFC.
\begin{figure}[t]
    \centering
    \vspace{0.5cm}
    \includegraphics[width=0.5\textwidth]{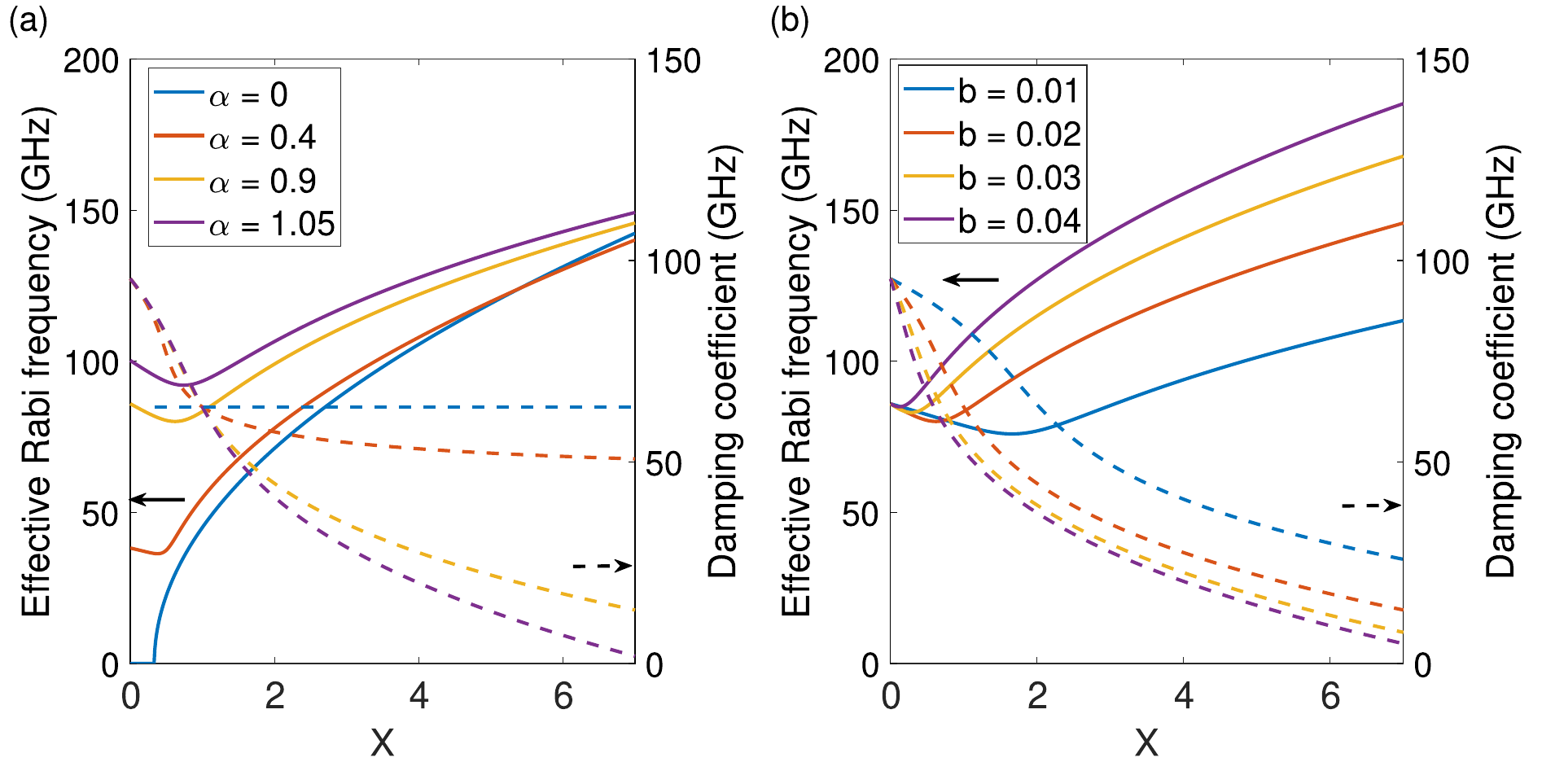}
    \caption{Effective Rabi frequency (solid) and damping coefficient (dashed) as a function of $X = |F|^{2}$ for different values of $\alpha$ (a) and $b$ (b). Solid and dashed curves of the same color indicate the same $\alpha$ (panel a) and $b$ (panel b). The ERF and the damping coefficient are calculated as specified in the Supplemental Material. In both figures $\Gamma=0.06$. In figure (a) $b=0.02$, while in (b) $\alpha=0.9$.}
    \label{fig_autov}
\end{figure}
With the final aim to describe the instabilities affecting a CW emission at threshold, we now study the conditions in which the medium coherently  behaves in an oscillatory fashion. We thus assume that the field $F$ is constant and monochromatic, i.e. the medium interacts with a CW at the reference frequency (the maximum of laser unsaturated gain). This results in a linear system governed by equations for \(P\), \(P^*\), and \(D\) the dynamics of which is described by the complex eigenvalues \(\beta\) of the system coefficient matrix $A$ (see Section III in the Supplementary materials). 
We highlight that an analogous procedure is adopted to estimate the Rabi frequency in a collection of two level atoms \cite{Nos,Torrey49}.
Since it turns out that for $\alpha\neq0$ two of these eigenvalues are complex conjugate and one is real, we identify the imaginary part of the complex solutions with  a characteristic frequency associated with the oscillations of the material variables ($P$ and $D$). We name Effective Rabi Frequency (ERF) this frequency in analogy with the Rabi Frequency of the two-level system.\\
In Fig.~\ref{fig_autov}(a), we plot the ERF as a function of $X = |F|^2$ for different values of the $\alpha$ factor (solid lines). For $\alpha \neq 0$, the curve exhibits a decreasing trend at low field intensity before transitioning into a portion that grows as $\sqrt{X}$.
As $\alpha$ diminishes, the decreasing portion progressively narrows until it vanishes completely at $\alpha = 0$, where, as expected, the curve follows a pure $\sqrt{X}$ trend, characteristic of the Rabi Frequency for two-level system case \cite{Nos} (see also Section III in the Supplementary materials). The dashed curves in Fig.~\ref{fig_autov}(a) show the $X$-dependence of the damping coefficient, given by the absolute value of the real part of $\beta$. For $X > 1$, the damping decreases with increasing $\alpha$, nearly vanishing for large $X$ and $\alpha = 1.05$—a typical value for QCLs \cite{Grillot16,SilvestriReview,PiccardoReview}—thus supporting the experimental observation of HFCs in these lasers. For $\alpha = 0$, in the region where $ERF = 0$, the damping coefficient is not plotted, as no oscillations occur and damping is therefore not well-defined.\\
In Fig.~\ref{fig_autov}(b), the ERF vs. $X$ curve (solid) is plotted for varying $b$. The segment corresponding to a decrease of the ERF with the field intensity becomes more extended for larger $b$, i.e. for a faster carrier dynamics, and disappears in the two-level limit, where the system becomes an overdamped oscillator ($ERF=0$).\\
Moreover, faster carrier dynamics, associated with larger $b$, lead to reduced damping (dashed curves in Fig.~\ref{fig_autov}(b)). This occurs also at low $X$ and agrees with the observation of HFCs near the laser threshold in mid-IR QCLs \cite{Mansuripur2016,Kazakov2017} differently from what happens in THz QCLs, which exhibit slower carrier dynamics. 
Furthermore, this evidence is in agreement with the experimental observation of Rabi flopping in QCLs and QD amplifiers that are characterized by a similar effective carriers lifetime \cite{Capua2014,Liu2010,Choi2010}. Consistently, in quantum well lasers—characterized by carrier lifetimes two to four orders of magnitude longer than those in QCLs— we verified that the damping is significantly higher (see Fig.~S5 in the Supplementary Materials) thus hindering the observation of Rabi oscillations and consequently HFCs in these devices.

From these evidences we may thus expect that, in analogy to what happens for the Rabi frequency in the RNGH instability in two level systems \cite{Nos}, the ERF could have a role in CW instability and eventually the emergence of OFCs. This was confirmed by performing the Linear Stability Analysis (LSA) of the CW solutions of Eqs.~\eqref{Eeq}-\eqref{Deq}, whose expressions are provided in Section I of the Supplementary Materials.\\
We observe that the CW corresponding to the minimum threshold, denoted as CW0 in the following, is that obtained by setting wavenumber $k=0$ in the steady state equations, i.e. the closest  to the gain peak (which is the reference frequency) \cite{Columbo2018}.

To study the stability of the CW0 solution against the growth of undamped modes, we determine the eigenvalues $\lambda_{n}$ of the Jacobian Matrix associated with the linear system for the perturbations that is obtained by seeking solutions to Eqs.~\eqref{Eeq}-\eqref{Deq} in the form
$F = (F_0 + \sum_n\delta F_{n}e^{-ik_{n}\eta} e^{\lambda_{n} t}) e^{-i k \eta + i \omega t}$,
$P = (P_0 + \sum_n\delta P_{n}e^{-ik_{n}\eta} e^{\lambda_{n} t}) e^{-i k \eta + i \omega t}$, and
$D = D_0 + \sum_n\delta D_{n}e^{-ik_{n}\eta} e^{\lambda_{n} t}$,
where $|\delta F_{n}|\ll|F_0|$, $|\delta P_{n}|\ll |P_0|$, $\delta D_{n}\ll D_0$ and $k_{n}=2\pi n/L$ are the Fourier modes. 
Details of the derivation are provided in Section II of the Supplementary Materials.
The parametric gain, defined as the maximum of the real part of $\lambda_{n}$, can then be calculated. This is evaluated at a frequency $\nu_\mathrm{n}=ck_{n}$ relative to the CW frequency and treated as a continuous variable  (see the Supplementary materials).\\
\begin{figure}[t]
    \begin{center}
    \includegraphics[width=0.5\textwidth]{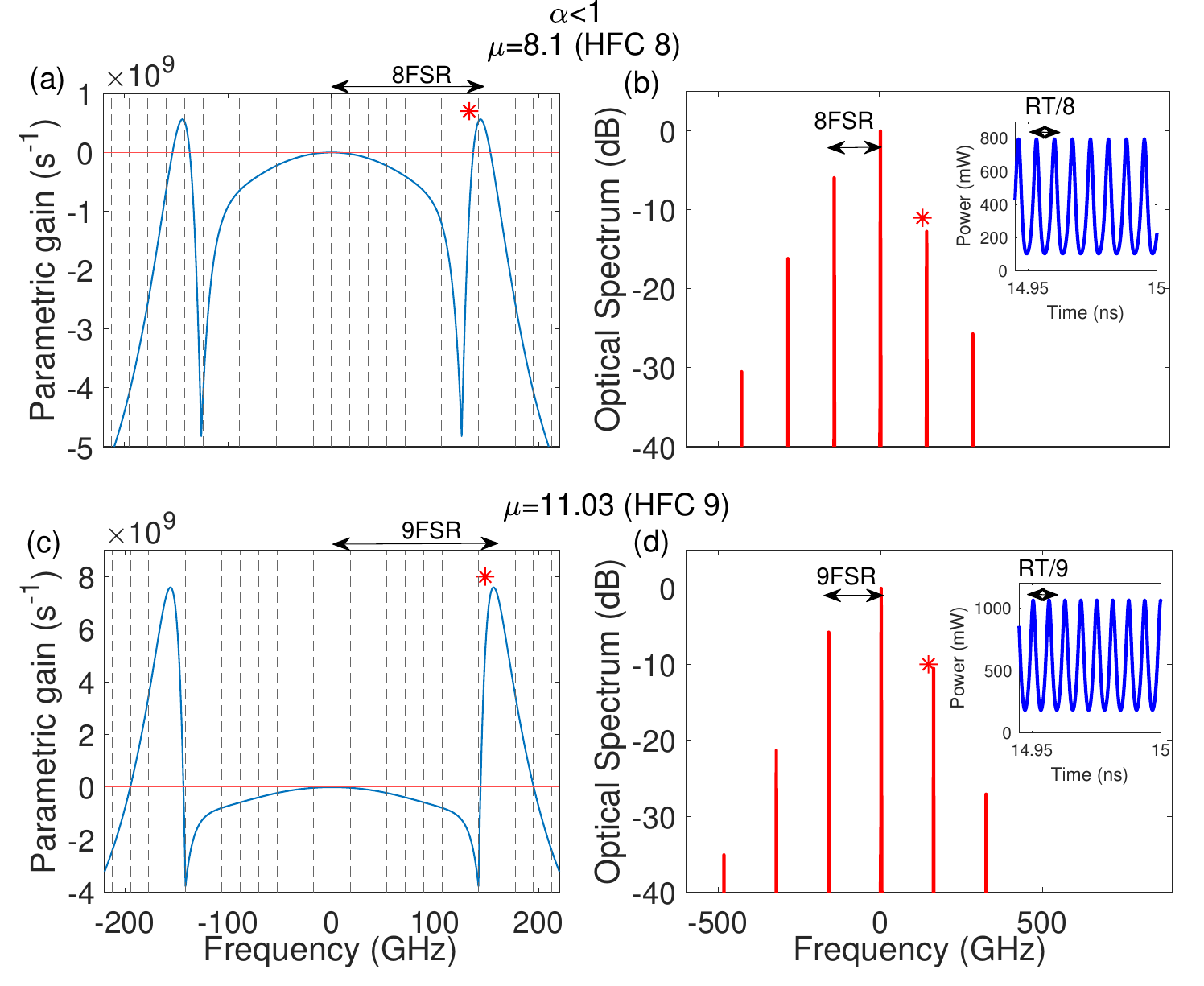}
       \end{center}
    \caption{Harmonic combs of different orders, obtained for different values of the pump parameter $\mu$ and for $\alpha=0.95$. The other parameters are $\Gamma=0.06$, $\mathrm{L}=4.7~$mm, $\sigma=1.6\times10^{-3}$, and $b=0.014$, with $\tau_e=7~\mathrm{ps}$ and $\tau_d=0.1~\mathrm{ps}$.  The left panels (figures (a), (c)) show the parametric gain, max$(Re(\lambda_{n}))$, as a function of frequency (blue curve), the cavity modes (black dashed lines). The right panels (figures (b), (d)) present the corresponding optical spectrum, obtained by numerically solving Eqs.~\eqref{Eeq}-\eqref{Deq}, and associated with HFCs of order 8 (top row) and 9 (bottom).  The effective Rabi frequency (red marker) is superimposed to both parametric gain and optical spectrum. The insets illustrate the numerically obtained temporal power trace of each HFC. RT: roundtrip time.}
    \label{pump_scan_alpha_minore1}
\end{figure}
\begin{figure}
\begin{center}
    \includegraphics[width=0.5\textwidth]{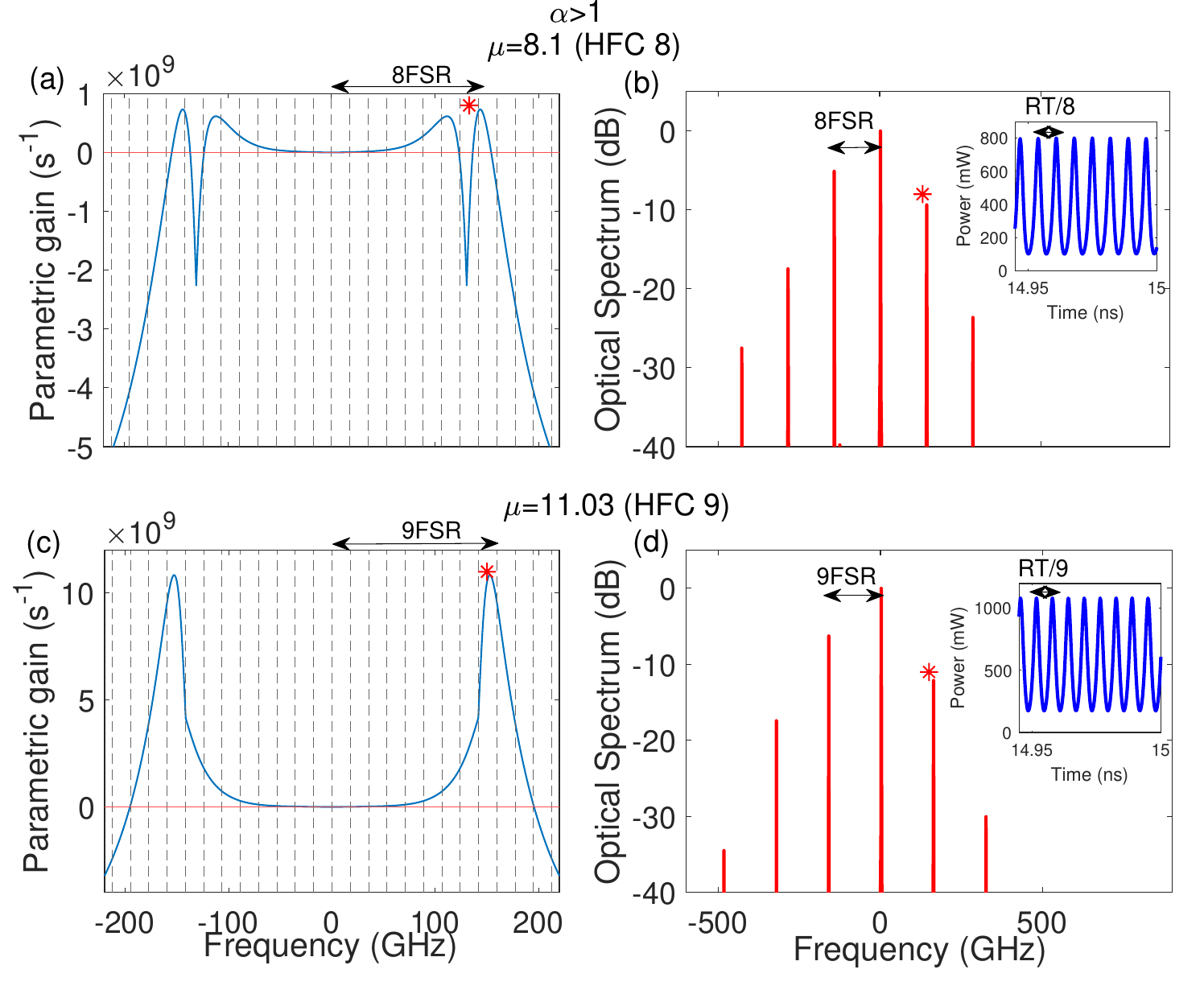}
       \end{center}
    \caption{Harmonic combs of different orders for $\alpha=1.01$. The parameters (except for $\alpha$) and the quantities plotted are the same as in Fig.~\ref{pump_scan_alpha_minore1}.}
    \label{pump_scan_alpha_maggiore1}
\end{figure}
Figure~\ref{pump_scan_alpha_minore1}(a) (blue curve) shows an example of the parametric gain numerically computed for typical parameters of a THz QCL and pump parameter $\mu=8.01$ (see figure caption). The frequency corresponding to the peak of the parametric gain coincides with the laser cavity mode (black dashed lines) spaced by 8 free spectral ranges (FSRs) from that of the CW0 ($k_\mathrm{0}=0$). Remarkably, the ERF (red marker) closely matches the parametric gain peak frequency. Solving numerically Eqs.\eqref{Eeq}-\eqref{Deq} for the same parameters as in Fig.~\ref{pump_scan_alpha_minore1}(a), shows that for the same parameters the CW is unstable and the system actually operates in a 8th-order harmonic comb regime (Fig.~\ref{pump_scan_alpha_minore1}(b)). 
Hence this suggests a connection between HFCs formation and the resonance between the ERF and one of the cavity modes.\\
Increasing $\mu$ to $11.03$ shifts the parametric gain peak frequency, which aligns with the cavity mode spaced 9 FSRs from the central frequency (Fig.~\ref{pump_scan_alpha_minore1}(c)). The ERF (red marker) also shifts, following the parametric gain peak. Consistent with the previous case, numerical simulations confirm the predictions of the LSA and the ERF, resulting in the numerical observation of a 9th-order HFC, as depicted in Fig.~\ref{pump_scan_alpha_minore1}(d).\\
The results presented in Fig.~\ref{pump_scan_alpha_minore1} were obtained for $\alpha < 1$. A similar behavior is also observed for $\alpha > 1$, although the shape of the parametric gain as a function of the frequency (and consequently the type of instability) differs from that in the $\alpha < 1$ case (see Fig.~\ref{pump_scan_alpha_maggiore1}). For $\alpha <1$, which is close to the two-level system case (formally corresponding to $\alpha=0$), the CW instability is analogous to the RNGH characterized by a balloon of unstable vectors $k_\mathrm{n}$ \cite{Nos,Columbo18Qdot} with a lower limit different from zero and with a maximum close to the ERF; instead for $\alpha > 1$, which represents a more realistic case for a QCL, especially in the mid-IR range \cite{Faist_book, Grillot16}, the CW instability is closer to the Benjamin-Feir or phase instability  since it corresponds to a balloon of unstable wave vectors $k_{n}$ with a lower limit equal to $k_{0}=0$ and with a maximum close to the ERF.
The BF instability is typical of the CGLE \cite{Aranson} that describes well a unidirectional ring QCL close to threshold \cite{NaturePiccardo, Columbo2021}.\\
In this instance (with $\alpha > 1$), we could observe that at lower values of $\mu$ 
compared to those in Fig.~\ref{pump_scan_alpha_maggiore1}, this Benjamin-Feir-like instability leads to the formation of OFCs associated with the presence of homoclons in the cavity (see Fig.~S4 in the Supplementary Material), as predicted and experimentally reported in \cite{NaturePiccardo}.\\
In some other cases we observe, for the same pump $\mu$, two distinct HFCs with different spacing and corresponding to the destabilization of different CWs, depending on how the pump parameter is scanned. Moreover, transitions between these solutions are also observed within a single simulation through a transient (see Fig.~S2 in Supplementary Material). A similar behavior has also been experimentally reported \cite{PiccardoHFCOptex}, though its interpretation remained unclear. In this work, we have chosen to present cases for values of $\mu$ where a single harmonic comb state is stably observed. 

\begin{figure}[t!]
    \begin{center}
    \includegraphics[width=0.5\textwidth]{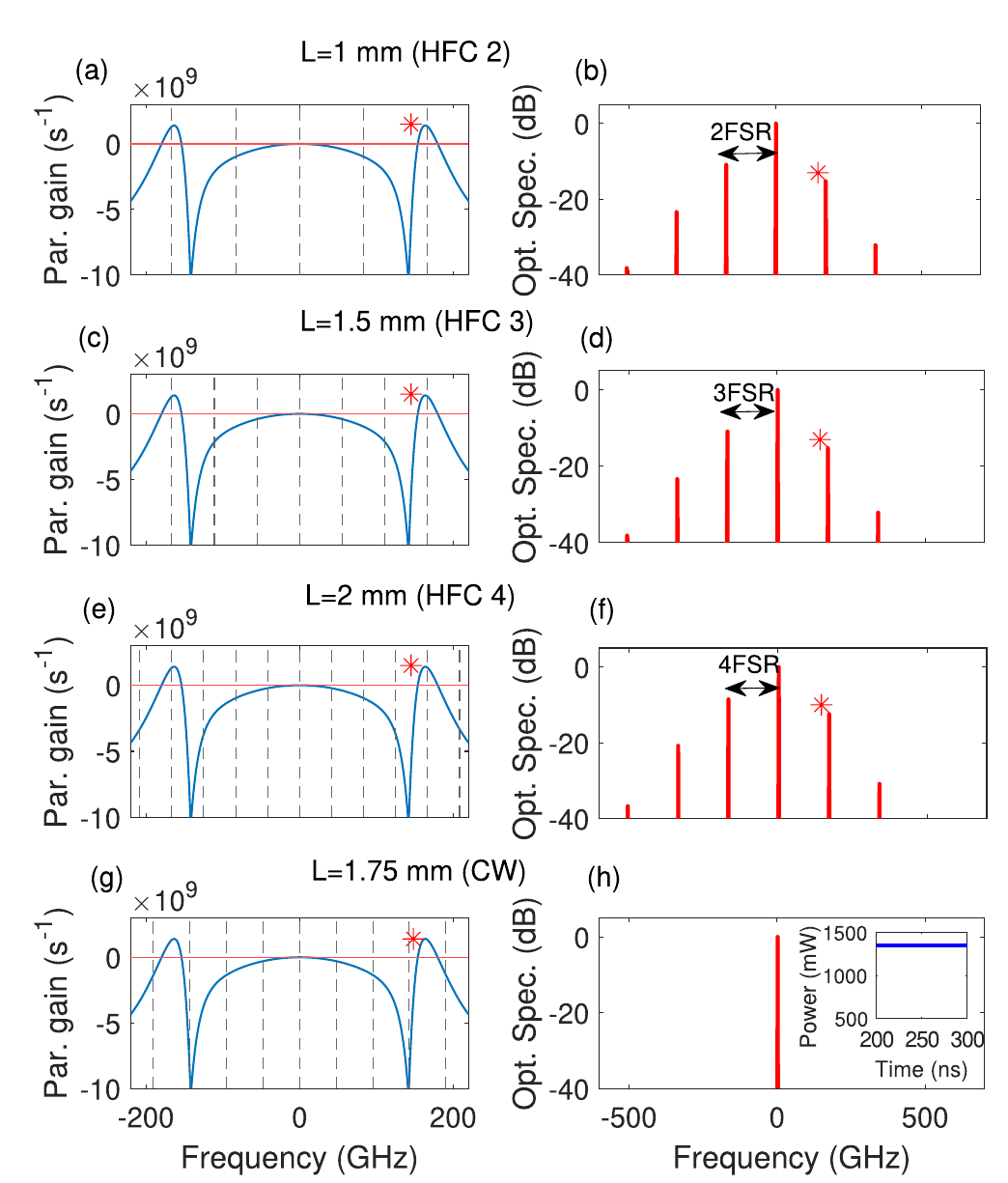}
       \end{center}
    \caption{Different regimes obtained by scanning the cavity length $L$ at fixed $\alpha=0.92$. For $L=1, 1.5, 2~$mm, HFCs of different order are reported in simulations, while for $L=1.75~$mm a CW is obtained. The other parameters are $\Gamma=0.06$, $\mu=7.9$, $\sigma=1.6\times10^{-3}$, and $b=0.02$, with $\tau_e=5~\mathrm{ps}$ and $\tau_d=0.1~\mathrm{ps}$. The left panels (figures (a), (c), (e), (g)) show the parametric gain as a function of the frequency. The right panels (figures (b), (d), (f), (h)) present the simulated optical spectrum. The inset in panel (h) illustrates the simulated temporal power trace for the CW regime. The power traces for the HFCs are trivially similar to those in previous figures, therefore they are omitted. The red markers indicate the estimated ERF.
    }
    \label{L_scan_alpha_minore1}
\end{figure}
To further validate our interpretation, we then investigate how the properties of the formation of HFCs as the cavity length varies, focusing first on the case of $\alpha < 1$ (Fig.~\ref{L_scan_alpha_minore1}). For $L=1$ mm and the other parameters set as reported in the figure caption, the parametric gain peak and the ERF are both located near the resonator mode at 2 FSRs from the CW frequency, consistent with numerical simulations showing a second-order HFC regime (Figs.\ref{L_scan_alpha_minore1}(a)-(b)). Increasing the cavity length to $L = 1.5$ mm reduces the free spectral range, shifting the parametric gain peak and ERF to align with a mode 3 FSRs from the central frequency. Simulations confirm the emergence of a third-order harmonic comb (Figs.\ref{L_scan_alpha_minore1}(c)-(d)). Similarly, for $L = 2$ mm, the gain peak and ERF coincide with a mode 4 FSRs from the central frequency, corresponding to a fourth-order harmonic comb (Figs.~\ref{L_scan_alpha_minore1}(e)-(f)). Remarkably, for $L = 1.75$ mm —intermediate between 1.5 mm and 2 mm—the peak of the parametric gain lies between two adjacent cavity modes (Fig.~\ref{L_scan_alpha_minore1}(g)), and no mode has a positive parametric gain. Consistently, numerical simulations reveal a stable CW regime instead of a harmonic comb (Fig.~\ref{L_scan_alpha_minore1}(h)). 
In this case as well, the ERF remains close to the peak of the parametric gain. This counterexample further corroborates the interpretation that HFCs formation arises from a resonance mechanism. 
Furthermore, we observe similar behavior by changing $L$ for $\alpha > 1$  albeit with the differences in the parametric gain lineshape mentioned above (see Section IV. in the Supplementary materials). \\
Therefore, further implications of a non-zero LEF  emerge, establishing a critical distinction between semiconductor and two-level system. Notably, the value of $\alpha$ dictates the shape of the parametric gain lines and the nature of the instability that underpins the formation of frequency combs.

The presented results highlight the significant predictive power of the LSA in forecasting HFC formation in a semiconductor laser. Furthermore, they allow to interpret the self-starting formation of HFC states as a resonance phenomenon that occurs when an intrinsic frequency of the system (the ERF) approaches one of the laser cavity modes. While this is reminiscent of the role played by the Rabi frequency in the onset of the RNGH instability in lasers with an active medium consisting of a collection of two-level systems \cite{Nos}, the theory we develop here applies to 
any semiconductor laser exhibiting multimode emission or frequency comb formation, with a significantly non-zero LEF.\\
Furthermore, understanding the physical mechanism behind the spontaneous formation of HFCs paves the way for a paradigm shift in the procedure for their generation and control. Whereas previous efforts have primarily focused on inducing and engineering these states using external interventions such as optical and electrical injection or a suitable reflectivity discontinuity or defects \cite{Dhillon1,Riccardi24,PiccardoOptical}, this work suggests alternative ways to tailor their properties. These include acting on bias conditions or on the QCL cavity design to tune the resonator mode spacing, for example, using mechanical \cite{Sumetsky10} or electro-optic \cite{Krasnokutska2019} methods.
Moreover, because of the suitability of QCLs for on-chip integration, the present evidence might accelerate the development of HFC-based photonic applications in spectroscopy and optical communications, that remained in their early stages due to the complexity of the experimental setups required so far for HFCs.

\bibliography{main}

\begin{thebibliography}{63}%
\makeatletter
\providecommand \@ifxundefined [1]{%
 \@ifx{#1\undefined}
}%
\providecommand \@ifnum [1]{%
 \ifnum #1\expandafter \@firstoftwo
 \else \expandafter \@secondoftwo
 \fi
}%
\providecommand \@ifx [1]{%
 \ifx #1\expandafter \@firstoftwo
 \else \expandafter \@secondoftwo
 \fi
}%
\providecommand \natexlab [1]{#1}%
\providecommand \enquote  [1]{``#1''}%
\providecommand \bibnamefont  [1]{#1}%
\providecommand \bibfnamefont [1]{#1}%
\providecommand \citenamefont [1]{#1}%
\providecommand \href@noop [0]{\@secondoftwo}%
\providecommand \href [0]{\begingroup \@sanitize@url \@href}%
\providecommand \@href[1]{\@@startlink{#1}\@@href}%
\providecommand \@@href[1]{\endgroup#1\@@endlink}%
\providecommand \@sanitize@url [0]{\catcode `\\12\catcode `\$12\catcode `\&12\catcode `\#12\catcode `\^12\catcode `\_12\catcode `\%12\relax}%
\providecommand \@@startlink[1]{}%
\providecommand \@@endlink[0]{}%
\providecommand \url  [0]{\begingroup\@sanitize@url \@url }%
\providecommand \@url [1]{\endgroup\@href {#1}{\urlprefix }}%
\providecommand \urlprefix  [0]{URL }%
\providecommand \Eprint [0]{\href }%
\providecommand \doibase [0]{https://doi.org/}%
\providecommand \selectlanguage [0]{\@gobble}%
\providecommand \bibinfo  [0]{\@secondoftwo}%
\providecommand \bibfield  [0]{\@secondoftwo}%
\providecommand \translation [1]{[#1]}%
\providecommand \BibitemOpen [0]{}%
\providecommand \bibitemStop [0]{}%
\providecommand \bibitemNoStop [0]{.\EOS\space}%
\providecommand \EOS [0]{\spacefactor3000\relax}%
\providecommand \BibitemShut  [1]{\csname bibitem#1\endcsname}%
\let\auto@bib@innerbib\@empty
\bibitem [{\citenamefont {H\"ansch}(2006)}]{Hansch_Nobel}%
  \BibitemOpen
  \bibfield  {author} {\bibinfo {author} {\bibfnamefont {T.~W.}\ \bibnamefont {H\"ansch}},\ }\bibfield  {title} {\bibinfo {title} {Nobel lecture: Passion for precision},\ }\href {https://doi.org/10.1103/RevModPhys.78.1297} {\bibfield  {journal} {\bibinfo  {journal} {Rev. Mod. Phys.}\ }\textbf {\bibinfo {volume} {78}},\ \bibinfo {pages} {1297} (\bibinfo {year} {2006})}\BibitemShut {NoStop}%
\bibitem [{\citenamefont {Weichman}\ \emph {et~al.}(2019)\citenamefont {Weichman}, \citenamefont {Changala}, \citenamefont {Ye}, \citenamefont {Chen}, \citenamefont {Yan},\ and\ \citenamefont {Picqué}}]{spectrev}%
  \BibitemOpen
  \bibfield  {author} {\bibinfo {author} {\bibfnamefont {M.~L.}\ \bibnamefont {Weichman}}, \bibinfo {author} {\bibfnamefont {P.~B.}\ \bibnamefont {Changala}}, \bibinfo {author} {\bibfnamefont {J.}~\bibnamefont {Ye}}, \bibinfo {author} {\bibfnamefont {Z.}~\bibnamefont {Chen}}, \bibinfo {author} {\bibfnamefont {M.}~\bibnamefont {Yan}},\ and\ \bibinfo {author} {\bibfnamefont {N.}~\bibnamefont {Picqué}},\ }\bibfield  {title} {\bibinfo {title} {Broadband molecular spectroscopy with optical frequency combs},\ }\href {https://doi.org/https://doi.org/10.1016/j.jms.2018.11.011} {\bibfield  {journal} {\bibinfo  {journal} {Journal of Molecular Spectroscopy}\ }\textbf {\bibinfo {volume} {355}},\ \bibinfo {pages} {66} (\bibinfo {year} {2019})}\BibitemShut {NoStop}%
\bibitem [{\citenamefont {Cing{\"o}z}\ \emph {et~al.}(2012)\citenamefont {Cing{\"o}z}, \citenamefont {Yost}, \citenamefont {Allison}, \citenamefont {Ruehl}, \citenamefont {Fermann}, \citenamefont {Hartl},\ and\ \citenamefont {Ye}}]{spect2012}%
  \BibitemOpen
  \bibfield  {author} {\bibinfo {author} {\bibfnamefont {A.}~\bibnamefont {Cing{\"o}z}}, \bibinfo {author} {\bibfnamefont {D.~C.}\ \bibnamefont {Yost}}, \bibinfo {author} {\bibfnamefont {T.~K.}\ \bibnamefont {Allison}}, \bibinfo {author} {\bibfnamefont {A.}~\bibnamefont {Ruehl}}, \bibinfo {author} {\bibfnamefont {M.~E.}\ \bibnamefont {Fermann}}, \bibinfo {author} {\bibfnamefont {I.}~\bibnamefont {Hartl}},\ and\ \bibinfo {author} {\bibfnamefont {J.}~\bibnamefont {Ye}},\ }\bibfield  {title} {\bibinfo {title} {Direct frequency comb spectroscopy in the extreme ultraviolet},\ }\href@noop {} {\bibfield  {journal} {\bibinfo  {journal} {Nature}\ }\textbf {\bibinfo {volume} {482}},\ \bibinfo {pages} {68} (\bibinfo {year} {2012})}\BibitemShut {NoStop}%
\bibitem [{\citenamefont {Fortier}\ and\ \citenamefont {Baumann}(2019)}]{Fortier2019}%
  \BibitemOpen
  \bibfield  {author} {\bibinfo {author} {\bibfnamefont {T.}~\bibnamefont {Fortier}}\ and\ \bibinfo {author} {\bibfnamefont {E.}~\bibnamefont {Baumann}},\ }\bibfield  {title} {\bibinfo {title} {20 years of developments in optical frequency comb technology and applications},\ }\href@noop {} {\bibfield  {journal} {\bibinfo  {journal} {Communications Physics}\ }\textbf {\bibinfo {volume} {2}},\ \bibinfo {pages} {153} (\bibinfo {year} {2019})}\BibitemShut {NoStop}%
\bibitem [{\citenamefont {Lomsadze}\ and\ \citenamefont {Cundiff}(2017)}]{spect2017}%
  \BibitemOpen
  \bibfield  {author} {\bibinfo {author} {\bibfnamefont {B.}~\bibnamefont {Lomsadze}}\ and\ \bibinfo {author} {\bibfnamefont {S.~T.}\ \bibnamefont {Cundiff}},\ }\bibfield  {title} {\bibinfo {title} {Frequency combs enable rapid and high-resolution multidimensional coherent spectroscopy},\ }\href@noop {} {\bibfield  {journal} {\bibinfo  {journal} {Science}\ }\textbf {\bibinfo {volume} {357}},\ \bibinfo {pages} {1389} (\bibinfo {year} {2017})}\BibitemShut {NoStop}%
\bibitem [{\citenamefont {Perego}\ and\ \citenamefont {Ellis}(2024)}]{perego2024optical}%
  \BibitemOpen
  \bibinfo {editor} {\bibfnamefont {A.~M.}\ \bibnamefont {Perego}}\ and\ \bibinfo {editor} {\bibfnamefont {A.}~\bibnamefont {Ellis}},\ eds.,\ \href {https://doi.org/10.1201/9781003427605} {\emph {\bibinfo {title} {Optical Frequency Combs: Trends in Sources and Applications}}},\ \bibinfo {edition} {1st}\ ed.\ (\bibinfo  {publisher} {CRC Press},\ \bibinfo {year} {2024})\BibitemShut {NoStop}%
\bibitem [{\citenamefont {Kippenberg}\ \emph {et~al.}(2011)\citenamefont {Kippenberg}, \citenamefont {Holzwarth},\ and\ \citenamefont {Diddams}}]{kippenbergrev}%
  \BibitemOpen
  \bibfield  {author} {\bibinfo {author} {\bibfnamefont {T.~J.}\ \bibnamefont {Kippenberg}}, \bibinfo {author} {\bibfnamefont {R.}~\bibnamefont {Holzwarth}},\ and\ \bibinfo {author} {\bibfnamefont {S.~A.}\ \bibnamefont {Diddams}},\ }\bibfield  {title} {\bibinfo {title} {Microresonator-based optical frequency combs},\ }\href@noop {} {\bibfield  {journal} {\bibinfo  {journal} {Science}\ }\textbf {\bibinfo {volume} {332}},\ \bibinfo {pages} {555} (\bibinfo {year} {2011})}\BibitemShut {NoStop}%
\bibitem [{\citenamefont {Faist}\ \emph {et~al.}(1994)\citenamefont {Faist}, \citenamefont {Capasso}, \citenamefont {Sivco}, \citenamefont {Sirtori}, \citenamefont {Hutchinson},\ and\ \citenamefont {Cho}}]{Faist_1994}%
  \BibitemOpen
  \bibfield  {author} {\bibinfo {author} {\bibfnamefont {J.}~\bibnamefont {Faist}}, \bibinfo {author} {\bibfnamefont {F.}~\bibnamefont {Capasso}}, \bibinfo {author} {\bibfnamefont {D.~L.}\ \bibnamefont {Sivco}}, \bibinfo {author} {\bibfnamefont {C.}~\bibnamefont {Sirtori}}, \bibinfo {author} {\bibfnamefont {A.~L.}\ \bibnamefont {Hutchinson}},\ and\ \bibinfo {author} {\bibfnamefont {A.~Y.}\ \bibnamefont {Cho}},\ }\bibfield  {title} {\bibinfo {title} {Quantum cascade laser},\ }\href@noop {} {\bibfield  {journal} {\bibinfo  {journal} {Science}\ }\textbf {\bibinfo {volume} {264}},\ \bibinfo {pages} {553} (\bibinfo {year} {1994})}\BibitemShut {NoStop}%
\bibitem [{\citenamefont {Lu}\ \emph {et~al.}(2009)\citenamefont {Lu}, \citenamefont {Liu}, \citenamefont {Poole}, \citenamefont {Raymond}, \citenamefont {Barrios}, \citenamefont {Poitras}, \citenamefont {Pakulski}, \citenamefont {Grant},\ and\ \citenamefont {Roy-Guay}}]{Lu09}%
  \BibitemOpen
  \bibfield  {author} {\bibinfo {author} {\bibfnamefont {Z.}~\bibnamefont {Lu}}, \bibinfo {author} {\bibfnamefont {J.}~\bibnamefont {Liu}}, \bibinfo {author} {\bibfnamefont {P.}~\bibnamefont {Poole}}, \bibinfo {author} {\bibfnamefont {S.}~\bibnamefont {Raymond}}, \bibinfo {author} {\bibfnamefont {P.}~\bibnamefont {Barrios}}, \bibinfo {author} {\bibfnamefont {D.}~\bibnamefont {Poitras}}, \bibinfo {author} {\bibfnamefont {G.}~\bibnamefont {Pakulski}}, \bibinfo {author} {\bibfnamefont {P.}~\bibnamefont {Grant}},\ and\ \bibinfo {author} {\bibfnamefont {D.}~\bibnamefont {Roy-Guay}},\ }\bibfield  {title} {\bibinfo {title} {An l-band monolithic inas/inp quantum dot mode-locked laser with femtosecond pulses},\ }\href {https://doi.org/10.1364/OE.17.013609} {\bibfield  {journal} {\bibinfo  {journal} {Opt. Express}\ }\textbf {\bibinfo {volume} {17}},\ \bibinfo {pages} {13609} (\bibinfo {year} {2009})}\BibitemShut {NoStop}%
\bibitem [{\citenamefont {Hugi}\ \emph {et~al.}(2012)\citenamefont {Hugi}, \citenamefont {Villares}, \citenamefont {Blaser}, \citenamefont {Liu},\ and\ \citenamefont {Faist}}]{Hugi2012}%
  \BibitemOpen
  \bibfield  {author} {\bibinfo {author} {\bibfnamefont {A.}~\bibnamefont {Hugi}}, \bibinfo {author} {\bibfnamefont {G.}~\bibnamefont {Villares}}, \bibinfo {author} {\bibfnamefont {S.}~\bibnamefont {Blaser}}, \bibinfo {author} {\bibfnamefont {H.~C.}\ \bibnamefont {Liu}},\ and\ \bibinfo {author} {\bibfnamefont {J.}~\bibnamefont {Faist}},\ }\bibfield  {title} {\bibinfo {title} {Mid-infrared frequency comb based on a quantum cascade laser},\ }\href {https://doi.org/10.1038/nature11620} {\bibfield  {journal} {\bibinfo  {journal} {Nature}\ }\textbf {\bibinfo {volume} {492}},\ \bibinfo {pages} {229} (\bibinfo {year} {2012})}\BibitemShut {NoStop}%
\bibitem [{\citenamefont {Faist}\ \emph {et~al.}(2016)\citenamefont {Faist}, \citenamefont {Villares}, \citenamefont {Scalari}, \citenamefont {Rösch}, \citenamefont {Bonzon}, \citenamefont {Hugi},\ and\ \citenamefont {Beck}}]{Faist_2016}%
  \BibitemOpen
  \bibfield  {author} {\bibinfo {author} {\bibfnamefont {J.}~\bibnamefont {Faist}}, \bibinfo {author} {\bibfnamefont {G.}~\bibnamefont {Villares}}, \bibinfo {author} {\bibfnamefont {G.}~\bibnamefont {Scalari}}, \bibinfo {author} {\bibfnamefont {M.}~\bibnamefont {Rösch}}, \bibinfo {author} {\bibfnamefont {C.}~\bibnamefont {Bonzon}}, \bibinfo {author} {\bibfnamefont {A.}~\bibnamefont {Hugi}},\ and\ \bibinfo {author} {\bibfnamefont {M.}~\bibnamefont {Beck}},\ }\bibfield  {title} {\bibinfo {title} {Quantum cascade laser frequency combs},\ }\href {https://doi.org/doi:10.1515/nanoph-2016-0015} {\bibfield  {journal} {\bibinfo  {journal} {Nanophotonics}\ }\textbf {\bibinfo {volume} {5}},\ \bibinfo {pages} {272} (\bibinfo {year} {2016})}\BibitemShut {NoStop}%
\bibitem [{\citenamefont {Kippenberg}\ \emph {et~al.}(2018)\citenamefont {Kippenberg}, \citenamefont {Gaeta}, \citenamefont {Lipson},\ and\ \citenamefont {Gorodetsky}}]{Kippenberg2018}%
  \BibitemOpen
  \bibfield  {author} {\bibinfo {author} {\bibfnamefont {T.~J.}\ \bibnamefont {Kippenberg}}, \bibinfo {author} {\bibfnamefont {A.~L.}\ \bibnamefont {Gaeta}}, \bibinfo {author} {\bibfnamefont {M.}~\bibnamefont {Lipson}},\ and\ \bibinfo {author} {\bibfnamefont {M.~L.}\ \bibnamefont {Gorodetsky}},\ }\bibfield  {title} {\bibinfo {title} {Dissipative {K}err solitons in optical microresonators},\ }\href@noop {} {\bibfield  {journal} {\bibinfo  {journal} {Science}\ }\textbf {\bibinfo {volume} {361}},\ \bibinfo {pages} {eaan8083} (\bibinfo {year} {2018})}\BibitemShut {NoStop}%
\bibitem [{\citenamefont {Bardella}\ \emph {et~al.}(2017)\citenamefont {Bardella}, \citenamefont {Columbo},\ and\ \citenamefont {Gioannini}}]{Bardella17}%
  \BibitemOpen
  \bibfield  {author} {\bibinfo {author} {\bibfnamefont {P.}~\bibnamefont {Bardella}}, \bibinfo {author} {\bibfnamefont {L.~L.}\ \bibnamefont {Columbo}},\ and\ \bibinfo {author} {\bibfnamefont {M.}~\bibnamefont {Gioannini}},\ }\bibfield  {title} {\bibinfo {title} {Self-generation of optical frequency comb in single section quantum dot {F}abry--{P}erot lasers: a theoretical study},\ }\href@noop {} {\bibfield  {journal} {\bibinfo  {journal} {Opt. Express}\ }\textbf {\bibinfo {volume} {25}},\ \bibinfo {pages} {26234} (\bibinfo {year} {2017})}\BibitemShut {NoStop}%
\bibitem [{\citenamefont {Faist}(2013)}]{Faist_book}%
  \BibitemOpen
  \bibfield  {author} {\bibinfo {author} {\bibfnamefont {J.}~\bibnamefont {Faist}},\ }\href@noop {} {\emph {\bibinfo {title} {Quantum Cascade Lasers}}}\ (\bibinfo  {publisher} {Oxford University Press},\ \bibinfo {year} {2013})\BibitemShut {NoStop}%
\bibitem [{\citenamefont {Burghoff}\ \emph {et~al.}(2014)\citenamefont {Burghoff}, \citenamefont {Kao}, \citenamefont {Han}, \citenamefont {Chan}, \citenamefont {Cai}, \citenamefont {Yang}, \citenamefont {Hayton}, \citenamefont {Gao}, \citenamefont {Reno},\ and\ \citenamefont {Hu}}]{Burghoff2014}%
  \BibitemOpen
  \bibfield  {author} {\bibinfo {author} {\bibfnamefont {D.}~\bibnamefont {Burghoff}}, \bibinfo {author} {\bibfnamefont {T.-Y.}\ \bibnamefont {Kao}}, \bibinfo {author} {\bibfnamefont {N.}~\bibnamefont {Han}}, \bibinfo {author} {\bibfnamefont {C.~W.~I.}\ \bibnamefont {Chan}}, \bibinfo {author} {\bibfnamefont {X.}~\bibnamefont {Cai}}, \bibinfo {author} {\bibfnamefont {Y.}~\bibnamefont {Yang}}, \bibinfo {author} {\bibfnamefont {D.~J.}\ \bibnamefont {Hayton}}, \bibinfo {author} {\bibfnamefont {J.-R.}\ \bibnamefont {Gao}}, \bibinfo {author} {\bibfnamefont {J.~L.}\ \bibnamefont {Reno}},\ and\ \bibinfo {author} {\bibfnamefont {Q.}~\bibnamefont {Hu}},\ }\bibfield  {title} {\bibinfo {title} {Terahertz laser frequency combs},\ }\href {https://doi.org/10.1038/nphoton.2014.85} {\bibfield  {journal} {\bibinfo  {journal} {Nature Photonics}\ }\textbf {\bibinfo {volume} {8}},\ \bibinfo {pages} {462} (\bibinfo {year} {2014})}\BibitemShut {NoStop}%
\bibitem [{\citenamefont {Villares}\ \emph {et~al.}(2014)\citenamefont {Villares}, \citenamefont {Hugi}, \citenamefont {Blaser},\ and\ \citenamefont {Faist}}]{Villares_2014}%
  \BibitemOpen
  \bibfield  {author} {\bibinfo {author} {\bibfnamefont {G.}~\bibnamefont {Villares}}, \bibinfo {author} {\bibfnamefont {A.}~\bibnamefont {Hugi}}, \bibinfo {author} {\bibfnamefont {S.}~\bibnamefont {Blaser}},\ and\ \bibinfo {author} {\bibfnamefont {J.}~\bibnamefont {Faist}},\ }\bibfield  {title} {\bibinfo {title} {Dual-comb spectroscopy based on quantum-cascade-laser frequency combs},\ }\href {https://doi.org/10.1038/ncomms6192} {\bibfield  {journal} {\bibinfo  {journal} {Nature Communications}\ }\textbf {\bibinfo {volume} {5}},\ \bibinfo {pages} {5192} (\bibinfo {year} {2014})}\BibitemShut {NoStop}%
\bibitem [{\citenamefont {Silvestri}\ \emph {et~al.}(2023{\natexlab{a}})\citenamefont {Silvestri}, \citenamefont {Qi}, \citenamefont {Taimre}, \citenamefont {Bertling},\ and\ \citenamefont {Rakić}}]{SilvestriReview}%
  \BibitemOpen
  \bibfield  {author} {\bibinfo {author} {\bibfnamefont {C.}~\bibnamefont {Silvestri}}, \bibinfo {author} {\bibfnamefont {X.}~\bibnamefont {Qi}}, \bibinfo {author} {\bibfnamefont {T.}~\bibnamefont {Taimre}}, \bibinfo {author} {\bibfnamefont {K.}~\bibnamefont {Bertling}},\ and\ \bibinfo {author} {\bibfnamefont {A.~D.}\ \bibnamefont {Rakić}},\ }\bibfield  {title} {\bibinfo {title} {Frequency combs in quantum cascade lasers: An overview of modeling and experiments},\ }\href@noop {} {\bibfield  {journal} {\bibinfo  {journal} {APL Photonics}\ }\textbf {\bibinfo {volume} {8}},\ \bibinfo {pages} {020902} (\bibinfo {year} {2023}{\natexlab{a}})}\BibitemShut {NoStop}%
\bibitem [{\citenamefont {Singleton}\ \emph {et~al.}(2018)\citenamefont {Singleton}, \citenamefont {Jouy}, \citenamefont {Beck},\ and\ \citenamefont {Faist}}]{Optica_Faist_18}%
  \BibitemOpen
  \bibfield  {author} {\bibinfo {author} {\bibfnamefont {M.}~\bibnamefont {Singleton}}, \bibinfo {author} {\bibfnamefont {P.}~\bibnamefont {Jouy}}, \bibinfo {author} {\bibfnamefont {M.}~\bibnamefont {Beck}},\ and\ \bibinfo {author} {\bibfnamefont {J.}~\bibnamefont {Faist}},\ }\bibfield  {title} {\bibinfo {title} {Evidence of linear chirp in mid-infrared quantum cascade lasers},\ }\href {https://doi.org/10.1364/OPTICA.5.000948} {\bibfield  {journal} {\bibinfo  {journal} {Optica}\ }\textbf {\bibinfo {volume} {5}},\ \bibinfo {pages} {948} (\bibinfo {year} {2018})}\BibitemShut {NoStop}%
\bibitem [{\citenamefont {Cappelli}\ \emph {et~al.}(2019)\citenamefont {Cappelli}, \citenamefont {Consolino}, \citenamefont {Campo}, \citenamefont {Galli}, \citenamefont {Mazzotti}, \citenamefont {Campa}, \citenamefont {Siciliani~de Cumis}, \citenamefont {Cancio~Pastor}, \citenamefont {Eramo}, \citenamefont {R{\"o}sch}, \citenamefont {Beck}, \citenamefont {Scalari}, \citenamefont {Faist}, \citenamefont {De~Natale},\ and\ \citenamefont {Bartalini}}]{Cappelli2019}%
  \BibitemOpen
  \bibfield  {author} {\bibinfo {author} {\bibfnamefont {F.}~\bibnamefont {Cappelli}}, \bibinfo {author} {\bibfnamefont {L.}~\bibnamefont {Consolino}}, \bibinfo {author} {\bibfnamefont {G.}~\bibnamefont {Campo}}, \bibinfo {author} {\bibfnamefont {I.}~\bibnamefont {Galli}}, \bibinfo {author} {\bibfnamefont {D.}~\bibnamefont {Mazzotti}}, \bibinfo {author} {\bibfnamefont {A.}~\bibnamefont {Campa}}, \bibinfo {author} {\bibfnamefont {M.}~\bibnamefont {Siciliani~de Cumis}}, \bibinfo {author} {\bibfnamefont {P.}~\bibnamefont {Cancio~Pastor}}, \bibinfo {author} {\bibfnamefont {R.}~\bibnamefont {Eramo}}, \bibinfo {author} {\bibfnamefont {M.}~\bibnamefont {R{\"o}sch}}, \bibinfo {author} {\bibfnamefont {M.}~\bibnamefont {Beck}}, \bibinfo {author} {\bibfnamefont {G.}~\bibnamefont {Scalari}}, \bibinfo {author} {\bibfnamefont {J.}~\bibnamefont {Faist}}, \bibinfo {author} {\bibfnamefont {P.}~\bibnamefont {De~Natale}},\ and\ \bibinfo {author} {\bibfnamefont {S.}~\bibnamefont {Bartalini}},\ }\bibfield  {title} {\bibinfo
  {title} {Retrieval of phase relation and emission profile of quantum cascade laser frequency combs},\ }\href@noop {} {\bibfield  {journal} {\bibinfo  {journal} {Nature Photonics}\ }\textbf {\bibinfo {volume} {13}},\ \bibinfo {pages} {562} (\bibinfo {year} {2019})}\BibitemShut {NoStop}%
\bibitem [{\citenamefont {Silvestri}\ \emph {et~al.}(2020)\citenamefont {Silvestri}, \citenamefont {Columbo}, \citenamefont {Brambilla},\ and\ \citenamefont {Gioannini}}]{Silvestri20}%
  \BibitemOpen
  \bibfield  {author} {\bibinfo {author} {\bibfnamefont {C.}~\bibnamefont {Silvestri}}, \bibinfo {author} {\bibfnamefont {L.~L.}\ \bibnamefont {Columbo}}, \bibinfo {author} {\bibfnamefont {M.}~\bibnamefont {Brambilla}},\ and\ \bibinfo {author} {\bibfnamefont {M.}~\bibnamefont {Gioannini}},\ }\bibfield  {title} {\bibinfo {title} {Coherent multi-mode dynamics in a quantum cascade laser: amplitude- and frequency-modulated optical frequency combs},\ }\href {https://doi.org/10.1364/OE.396481} {\bibfield  {journal} {\bibinfo  {journal} {Opt. Express}\ }\textbf {\bibinfo {volume} {28}},\ \bibinfo {pages} {23846} (\bibinfo {year} {2020})}\BibitemShut {NoStop}%
\bibitem [{\citenamefont {Opa\ifmmode~\check{c}\else \v{c}\fi{}ak}\ and\ \citenamefont {Schwarz}(2019)}]{Opacak2019}%
  \BibitemOpen
  \bibfield  {author} {\bibinfo {author} {\bibfnamefont {N.}~\bibnamefont {Opa\ifmmode~\check{c}\else \v{c}\fi{}ak}}\ and\ \bibinfo {author} {\bibfnamefont {B.}~\bibnamefont {Schwarz}},\ }\bibfield  {title} {\bibinfo {title} {Theory of frequency-modulated combs in lasers with spatial hole burning, dispersion, and {K}err nonlinearity},\ }\href {https://doi.org/10.1103/PhysRevLett.123.243902} {\bibfield  {journal} {\bibinfo  {journal} {Phys. Rev. Lett.}\ }\textbf {\bibinfo {volume} {123}},\ \bibinfo {pages} {243902} (\bibinfo {year} {2019})}\BibitemShut {NoStop}%
\bibitem [{\citenamefont {Burghoff}(2020)}]{Burghoff20}%
  \BibitemOpen
  \bibfield  {author} {\bibinfo {author} {\bibfnamefont {D.}~\bibnamefont {Burghoff}},\ }\bibfield  {title} {\bibinfo {title} {Unraveling the origin of frequency modulated combs using active cavity mean-field theory},\ }\href {http://opg.optica.org/optica/abstract.cfm?URI=optica-7-12-1781} {\bibfield  {journal} {\bibinfo  {journal} {Optica}\ }\textbf {\bibinfo {volume} {7}},\ \bibinfo {pages} {1781} (\bibinfo {year} {2020})}\BibitemShut {NoStop}%
\bibitem [{\citenamefont {Piccardo}\ \emph {et~al.}(2019)\citenamefont {Piccardo}, \citenamefont {Chevalier}, \citenamefont {Schwarz}, \citenamefont {Kazakov}, \citenamefont {Wang}, \citenamefont {Belyanin},\ and\ \citenamefont {Capasso}}]{Piccardo_PRL}%
  \BibitemOpen
  \bibfield  {author} {\bibinfo {author} {\bibfnamefont {M.}~\bibnamefont {Piccardo}}, \bibinfo {author} {\bibfnamefont {P.}~\bibnamefont {Chevalier}}, \bibinfo {author} {\bibfnamefont {B.}~\bibnamefont {Schwarz}}, \bibinfo {author} {\bibfnamefont {D.}~\bibnamefont {Kazakov}}, \bibinfo {author} {\bibfnamefont {Y.}~\bibnamefont {Wang}}, \bibinfo {author} {\bibfnamefont {A.}~\bibnamefont {Belyanin}},\ and\ \bibinfo {author} {\bibfnamefont {F.}~\bibnamefont {Capasso}},\ }\bibfield  {title} {\bibinfo {title} {Frequency-modulated combs obey a variational principle},\ }\href {https://doi.org/10.1103/PhysRevLett.122.253901} {\bibfield  {journal} {\bibinfo  {journal} {Phys. Rev. Lett.}\ }\textbf {\bibinfo {volume} {122}},\ \bibinfo {pages} {253901} (\bibinfo {year} {2019})}\BibitemShut {NoStop}%
\bibitem [{\citenamefont {Columbo}\ \emph {et~al.}(2018{\natexlab{a}})\citenamefont {Columbo}, \citenamefont {Barbieri}, \citenamefont {Sirtori},\ and\ \citenamefont {Brambilla}}]{Columbo2018}%
  \BibitemOpen
  \bibfield  {author} {\bibinfo {author} {\bibfnamefont {L.~L.}\ \bibnamefont {Columbo}}, \bibinfo {author} {\bibfnamefont {S.}~\bibnamefont {Barbieri}}, \bibinfo {author} {\bibfnamefont {C.}~\bibnamefont {Sirtori}},\ and\ \bibinfo {author} {\bibfnamefont {M.}~\bibnamefont {Brambilla}},\ }\bibfield  {title} {\bibinfo {title} {Dynamics of a broad-band quantum cascade laser: from chaos to coherent dynamics and mode-locking},\ }\href {http://opg.optica.org/oe/abstract.cfm?URI=oe-26-3-2829} {\bibfield  {journal} {\bibinfo  {journal} {Opt. Express}\ }\textbf {\bibinfo {volume} {26}},\ \bibinfo {pages} {2829} (\bibinfo {year} {2018}{\natexlab{a}})}\BibitemShut {NoStop}%
\bibitem [{\citenamefont {Piccardo}\ \emph {et~al.}(2020)\citenamefont {Piccardo}, \citenamefont {Schwarz}, \citenamefont {Kazakov}, \citenamefont {Beiser}, \citenamefont {Opa{\v{c}}ak}, \citenamefont {Wang}, \citenamefont {Jha}, \citenamefont {Hillbrand}, \citenamefont {Tamagnone}, \citenamefont {Chen}, \citenamefont {Zhu}, \citenamefont {Columbo}, \citenamefont {Belyanin},\ and\ \citenamefont {Capasso}}]{NaturePiccardo}%
  \BibitemOpen
  \bibfield  {author} {\bibinfo {author} {\bibfnamefont {M.}~\bibnamefont {Piccardo}}, \bibinfo {author} {\bibfnamefont {B.}~\bibnamefont {Schwarz}}, \bibinfo {author} {\bibfnamefont {D.}~\bibnamefont {Kazakov}}, \bibinfo {author} {\bibfnamefont {M.}~\bibnamefont {Beiser}}, \bibinfo {author} {\bibfnamefont {N.}~\bibnamefont {Opa{\v{c}}ak}}, \bibinfo {author} {\bibfnamefont {Y.}~\bibnamefont {Wang}}, \bibinfo {author} {\bibfnamefont {S.}~\bibnamefont {Jha}}, \bibinfo {author} {\bibfnamefont {J.}~\bibnamefont {Hillbrand}}, \bibinfo {author} {\bibfnamefont {M.}~\bibnamefont {Tamagnone}}, \bibinfo {author} {\bibfnamefont {W.~T.}\ \bibnamefont {Chen}}, \bibinfo {author} {\bibfnamefont {A.~Y.}\ \bibnamefont {Zhu}}, \bibinfo {author} {\bibfnamefont {L.~L.}\ \bibnamefont {Columbo}}, \bibinfo {author} {\bibfnamefont {A.}~\bibnamefont {Belyanin}},\ and\ \bibinfo {author} {\bibfnamefont {F.}~\bibnamefont {Capasso}},\ }\bibfield  {title} {\bibinfo {title} {Frequency combs induced by phase turbulence},\ }\href
  {https://doi.org/10.1038/s41586-020-2386-6} {\bibfield  {journal} {\bibinfo  {journal} {Nature}\ }\textbf {\bibinfo {volume} {582}},\ \bibinfo {pages} {360} (\bibinfo {year} {2020})}\BibitemShut {NoStop}%
\bibitem [{\citenamefont {Meng}\ \emph {et~al.}(2020)\citenamefont {Meng}, \citenamefont {Singleton}, \citenamefont {Shahmohammadi}, \citenamefont {Kapsalidis}, \citenamefont {Wang}, \citenamefont {Beck},\ and\ \citenamefont {Faist}}]{Bomeng1}%
  \BibitemOpen
  \bibfield  {author} {\bibinfo {author} {\bibfnamefont {B.}~\bibnamefont {Meng}}, \bibinfo {author} {\bibfnamefont {M.}~\bibnamefont {Singleton}}, \bibinfo {author} {\bibfnamefont {M.}~\bibnamefont {Shahmohammadi}}, \bibinfo {author} {\bibfnamefont {F.}~\bibnamefont {Kapsalidis}}, \bibinfo {author} {\bibfnamefont {R.}~\bibnamefont {Wang}}, \bibinfo {author} {\bibfnamefont {M.}~\bibnamefont {Beck}},\ and\ \bibinfo {author} {\bibfnamefont {J.}~\bibnamefont {Faist}},\ }\bibfield  {title} {\bibinfo {title} {Mid-infrared frequency comb from a ring quantum cascade laser},\ }\href {https://doi.org/10.1364/OPTICA.377755} {\bibfield  {journal} {\bibinfo  {journal} {Optica}\ }\textbf {\bibinfo {volume} {7}},\ \bibinfo {pages} {162} (\bibinfo {year} {2020})}\BibitemShut {NoStop}%
\bibitem [{\citenamefont {Meng}\ \emph {et~al.}(2022)\citenamefont {Meng}, \citenamefont {Singleton}, \citenamefont {Hillbrand}, \citenamefont {Francki{\'e}}, \citenamefont {Beck},\ and\ \citenamefont {Faist}}]{Bomeng2}%
  \BibitemOpen
  \bibfield  {author} {\bibinfo {author} {\bibfnamefont {B.}~\bibnamefont {Meng}}, \bibinfo {author} {\bibfnamefont {M.}~\bibnamefont {Singleton}}, \bibinfo {author} {\bibfnamefont {J.}~\bibnamefont {Hillbrand}}, \bibinfo {author} {\bibfnamefont {M.}~\bibnamefont {Francki{\'e}}}, \bibinfo {author} {\bibfnamefont {M.}~\bibnamefont {Beck}},\ and\ \bibinfo {author} {\bibfnamefont {J.}~\bibnamefont {Faist}},\ }\bibfield  {title} {\bibinfo {title} {Dissipative {K}err solitons in semiconductor ring lasers},\ }\href {https://doi.org/10.1038/s41566-021-00927-3} {\bibfield  {journal} {\bibinfo  {journal} {Nature Photonics}\ }\textbf {\bibinfo {volume} {16}},\ \bibinfo {pages} {142} (\bibinfo {year} {2022})}\BibitemShut {NoStop}%
\bibitem [{\citenamefont {Columbo}\ \emph {et~al.}(2021)\citenamefont {Columbo}, \citenamefont {Piccardo}, \citenamefont {Prati}, \citenamefont {Lugiato}, \citenamefont {Brambilla}, \citenamefont {Gatti}, \citenamefont {Silvestri}, \citenamefont {Gioannini}, \citenamefont {Opa\ifmmode~\check{c}\else \v{c}\fi{}ak}, \citenamefont {Schwarz},\ and\ \citenamefont {Capasso}}]{Columbo2021}%
  \BibitemOpen
  \bibfield  {author} {\bibinfo {author} {\bibfnamefont {L.}~\bibnamefont {Columbo}}, \bibinfo {author} {\bibfnamefont {M.}~\bibnamefont {Piccardo}}, \bibinfo {author} {\bibfnamefont {F.}~\bibnamefont {Prati}}, \bibinfo {author} {\bibfnamefont {L.~A.}\ \bibnamefont {Lugiato}}, \bibinfo {author} {\bibfnamefont {M.}~\bibnamefont {Brambilla}}, \bibinfo {author} {\bibfnamefont {A.}~\bibnamefont {Gatti}}, \bibinfo {author} {\bibfnamefont {C.}~\bibnamefont {Silvestri}}, \bibinfo {author} {\bibfnamefont {M.}~\bibnamefont {Gioannini}}, \bibinfo {author} {\bibfnamefont {N.}~\bibnamefont {Opa\ifmmode~\check{c}\else \v{c}\fi{}ak}}, \bibinfo {author} {\bibfnamefont {B.}~\bibnamefont {Schwarz}},\ and\ \bibinfo {author} {\bibfnamefont {F.}~\bibnamefont {Capasso}},\ }\bibfield  {title} {\bibinfo {title} {Unifying frequency combs in active and passive cavities: Temporal solitons in externally driven ring lasers},\ }\href {https://doi.org/10.1103/PhysRevLett.126.173903} {\bibfield  {journal} {\bibinfo  {journal} {Phys. Rev.
  Lett.}\ }\textbf {\bibinfo {volume} {126}},\ \bibinfo {pages} {173903} (\bibinfo {year} {2021})}\BibitemShut {NoStop}%
\bibitem [{\citenamefont {Opačak}\ \emph {et~al.}(2024)\citenamefont {Opačak}, \citenamefont {Kazakov}, \citenamefont {Columbo}, \citenamefont {Beiser}, \citenamefont {Letsou}, \citenamefont {Pilat}, \citenamefont {Brambilla}, \citenamefont {Prati}, \citenamefont {Piccardo}, \citenamefont {Capasso},\ and\ \citenamefont {Schwarz}}]{NB2024}%
  \BibitemOpen
  \bibfield  {author} {\bibinfo {author} {\bibfnamefont {N.}~\bibnamefont {Opačak}}, \bibinfo {author} {\bibfnamefont {D.}~\bibnamefont {Kazakov}}, \bibinfo {author} {\bibfnamefont {L.}~\bibnamefont {Columbo}}, \bibinfo {author} {\bibfnamefont {M.}~\bibnamefont {Beiser}}, \bibinfo {author} {\bibfnamefont {T.~P.}\ \bibnamefont {Letsou}}, \bibinfo {author} {\bibfnamefont {F.}~\bibnamefont {Pilat}}, \bibinfo {author} {\bibfnamefont {M.}~\bibnamefont {Brambilla}}, \bibinfo {author} {\bibfnamefont {F.}~\bibnamefont {Prati}}, \bibinfo {author} {\bibfnamefont {M.}~\bibnamefont {Piccardo}}, \bibinfo {author} {\bibfnamefont {F.}~\bibnamefont {Capasso}},\ and\ \bibinfo {author} {\bibfnamefont {B.}~\bibnamefont {Schwarz}},\ }\bibfield  {title} {\bibinfo {title} {Nozaki–{B}ekki solitons in semiconductor lasers},\ }\href@noop {} {\bibfield  {journal} {\bibinfo  {journal} {Nature}\ }\textbf {\bibinfo {volume} {625}},\ \bibinfo {pages} {685} (\bibinfo {year} {Jan 2024})}\BibitemShut {NoStop}%
\bibitem [{\citenamefont {Pasquazi}\ \emph {et~al.}(2018)\citenamefont {Pasquazi}, \citenamefont {Peccianti}, \citenamefont {Razzari}, \citenamefont {Moss}, \citenamefont {Coen}, \citenamefont {Erkintalo}, \citenamefont {Chembo}, \citenamefont {Hansson}, \citenamefont {Wabnitz}, \citenamefont {Del’Haye}, \citenamefont {Xue}, \citenamefont {Weiner},\ and\ \citenamefont {Morandotti}}]{Pasquazi2018}%
  \BibitemOpen
  \bibfield  {author} {\bibinfo {author} {\bibfnamefont {A.}~\bibnamefont {Pasquazi}}, \bibinfo {author} {\bibfnamefont {M.}~\bibnamefont {Peccianti}}, \bibinfo {author} {\bibfnamefont {L.}~\bibnamefont {Razzari}}, \bibinfo {author} {\bibfnamefont {D.~J.}\ \bibnamefont {Moss}}, \bibinfo {author} {\bibfnamefont {S.}~\bibnamefont {Coen}}, \bibinfo {author} {\bibfnamefont {M.}~\bibnamefont {Erkintalo}}, \bibinfo {author} {\bibfnamefont {Y.~K.}\ \bibnamefont {Chembo}}, \bibinfo {author} {\bibfnamefont {T.}~\bibnamefont {Hansson}}, \bibinfo {author} {\bibfnamefont {S.}~\bibnamefont {Wabnitz}}, \bibinfo {author} {\bibfnamefont {P.}~\bibnamefont {Del’Haye}}, \bibinfo {author} {\bibfnamefont {X.}~\bibnamefont {Xue}}, \bibinfo {author} {\bibfnamefont {A.~M.}\ \bibnamefont {Weiner}},\ and\ \bibinfo {author} {\bibfnamefont {R.}~\bibnamefont {Morandotti}},\ }\bibfield  {title} {\bibinfo {title} {Micro-combs: A novel generation of optical sources},\ }\href {https://doi.org/https://doi.org/10.1016/j.physrep.2017.08.004}
  {\bibfield  {journal} {\bibinfo  {journal} {Physics Reports}\ }\textbf {\bibinfo {volume} {729}},\ \bibinfo {pages} {1} (\bibinfo {year} {2018})},\ \bibinfo {note} {micro-combs: A novel generation of optical sources}\BibitemShut {NoStop}%
\bibitem [{\citenamefont {Mansuripur}\ \emph {et~al.}(2016)\citenamefont {Mansuripur}, \citenamefont {Vernet}, \citenamefont {Chevalier}, \citenamefont {Aoust}, \citenamefont {Schwarz}, \citenamefont {Xie}, \citenamefont {Caneau}, \citenamefont {Lascola}, \citenamefont {Zah}, \citenamefont {Caffey}, \citenamefont {Day}, \citenamefont {Missaggia}, \citenamefont {Connors}, \citenamefont {Wang}, \citenamefont {Belyanin},\ and\ \citenamefont {Capasso}}]{Mansuripur2016}%
  \BibitemOpen
  \bibfield  {author} {\bibinfo {author} {\bibfnamefont {T.~S.}\ \bibnamefont {Mansuripur}}, \bibinfo {author} {\bibfnamefont {C.}~\bibnamefont {Vernet}}, \bibinfo {author} {\bibfnamefont {P.}~\bibnamefont {Chevalier}}, \bibinfo {author} {\bibfnamefont {G.}~\bibnamefont {Aoust}}, \bibinfo {author} {\bibfnamefont {B.}~\bibnamefont {Schwarz}}, \bibinfo {author} {\bibfnamefont {F.}~\bibnamefont {Xie}}, \bibinfo {author} {\bibfnamefont {C.}~\bibnamefont {Caneau}}, \bibinfo {author} {\bibfnamefont {K.}~\bibnamefont {Lascola}}, \bibinfo {author} {\bibfnamefont {C.}~\bibnamefont {Zah}}, \bibinfo {author} {\bibfnamefont {D.~P.}\ \bibnamefont {Caffey}}, \bibinfo {author} {\bibfnamefont {T.}~\bibnamefont {Day}}, \bibinfo {author} {\bibfnamefont {L.~J.}\ \bibnamefont {Missaggia}}, \bibinfo {author} {\bibfnamefont {M.~K.}\ \bibnamefont {Connors}}, \bibinfo {author} {\bibfnamefont {C.~A.}\ \bibnamefont {Wang}}, \bibinfo {author} {\bibfnamefont {A.}~\bibnamefont {Belyanin}},\ and\ \bibinfo {author} {\bibfnamefont
  {F.}~\bibnamefont {Capasso}},\ }\bibfield  {title} {\bibinfo {title} {Single-mode instability in standing-wave lasers: The quantum cascade laser as a self-pumped parametric oscillator},\ }\href {https://doi.org/10.1103/PhysRevA.94.063807} {\bibfield  {journal} {\bibinfo  {journal} {Phys. Rev. A}\ }\textbf {\bibinfo {volume} {94}},\ \bibinfo {pages} {063807} (\bibinfo {year} {2016})}\BibitemShut {NoStop}%
\bibitem [{\citenamefont {Kazakov}\ \emph {et~al.}(2017)\citenamefont {Kazakov}, \citenamefont {Piccardo}, \citenamefont {Wang}, \citenamefont {Chevalier}, \citenamefont {Mansuripur}, \citenamefont {Xie}, \citenamefont {Zah}, \citenamefont {Lascola}, \citenamefont {Belyanin},\ and\ \citenamefont {Capasso}}]{Kazakov2017}%
  \BibitemOpen
  \bibfield  {author} {\bibinfo {author} {\bibfnamefont {D.}~\bibnamefont {Kazakov}}, \bibinfo {author} {\bibfnamefont {M.}~\bibnamefont {Piccardo}}, \bibinfo {author} {\bibfnamefont {Y.}~\bibnamefont {Wang}}, \bibinfo {author} {\bibfnamefont {P.}~\bibnamefont {Chevalier}}, \bibinfo {author} {\bibfnamefont {T.~S.}\ \bibnamefont {Mansuripur}}, \bibinfo {author} {\bibfnamefont {F.}~\bibnamefont {Xie}}, \bibinfo {author} {\bibfnamefont {C.-e.}\ \bibnamefont {Zah}}, \bibinfo {author} {\bibfnamefont {K.}~\bibnamefont {Lascola}}, \bibinfo {author} {\bibfnamefont {A.}~\bibnamefont {Belyanin}},\ and\ \bibinfo {author} {\bibfnamefont {F.}~\bibnamefont {Capasso}},\ }\bibfield  {title} {\bibinfo {title} {Self-starting harmonic frequency comb generation in a quantum cascade laser},\ }\href@noop {} {\bibfield  {journal} {\bibinfo  {journal} {Nature Photonics}\ }\textbf {\bibinfo {volume} {11}},\ \bibinfo {pages} {789} (\bibinfo {year} {2017})}\BibitemShut {NoStop}%
\bibitem [{\citenamefont {Wang}\ \emph {et~al.}(2020{\natexlab{a}})\citenamefont {Wang}, \citenamefont {Pistore}, \citenamefont {Riesch}, \citenamefont {Nong}, \citenamefont {Vigneron}, \citenamefont {Colombelli}, \citenamefont {Parillaud}, \citenamefont {Mangeney}, \citenamefont {Tignon}, \citenamefont {Jirauschek},\ and\ \citenamefont {Dhillon}}]{Wang2020}%
  \BibitemOpen
  \bibfield  {author} {\bibinfo {author} {\bibfnamefont {F.}~\bibnamefont {Wang}}, \bibinfo {author} {\bibfnamefont {V.}~\bibnamefont {Pistore}}, \bibinfo {author} {\bibfnamefont {M.}~\bibnamefont {Riesch}}, \bibinfo {author} {\bibfnamefont {H.}~\bibnamefont {Nong}}, \bibinfo {author} {\bibfnamefont {P.-B.}\ \bibnamefont {Vigneron}}, \bibinfo {author} {\bibfnamefont {R.}~\bibnamefont {Colombelli}}, \bibinfo {author} {\bibfnamefont {O.}~\bibnamefont {Parillaud}}, \bibinfo {author} {\bibfnamefont {J.}~\bibnamefont {Mangeney}}, \bibinfo {author} {\bibfnamefont {J.}~\bibnamefont {Tignon}}, \bibinfo {author} {\bibfnamefont {C.}~\bibnamefont {Jirauschek}},\ and\ \bibinfo {author} {\bibfnamefont {S.~S.}\ \bibnamefont {Dhillon}},\ }\bibfield  {title} {\bibinfo {title} {Ultrafast response of harmonic modelocked thz lasers},\ }\href {https://doi.org/10.1038/s41377-020-0288-x} {\bibfield  {journal} {\bibinfo  {journal} {Light: Science {\&} Applications}\ }\textbf {\bibinfo {volume} {9}},\ \bibinfo {pages} {51} (\bibinfo
  {year} {2020}{\natexlab{a}})}\BibitemShut {NoStop}%
\bibitem [{\citenamefont {Forrer}\ \emph {et~al.}(2021)\citenamefont {Forrer}, \citenamefont {Wang}, \citenamefont {Beck}, \citenamefont {Belyanin}, \citenamefont {Faist},\ and\ \citenamefont {Scalari}}]{ForrerHFC}%
  \BibitemOpen
  \bibfield  {author} {\bibinfo {author} {\bibfnamefont {A.}~\bibnamefont {Forrer}}, \bibinfo {author} {\bibfnamefont {Y.}~\bibnamefont {Wang}}, \bibinfo {author} {\bibfnamefont {M.}~\bibnamefont {Beck}}, \bibinfo {author} {\bibfnamefont {A.}~\bibnamefont {Belyanin}}, \bibinfo {author} {\bibfnamefont {J.}~\bibnamefont {Faist}},\ and\ \bibinfo {author} {\bibfnamefont {G.}~\bibnamefont {Scalari}},\ }\bibfield  {title} {\bibinfo {title} {Self-starting harmonic comb emission in {TH}z quantum cascade lasers},\ }\href@noop {} {\bibfield  {journal} {\bibinfo  {journal} {Applied Physics Letters}\ }\textbf {\bibinfo {volume} {118}},\ \bibinfo {pages} {131112} (\bibinfo {year} {2021})}\BibitemShut {NoStop}%
\bibitem [{\citenamefont {Jaidl}\ \emph {et~al.}(2021)\citenamefont {Jaidl}, \citenamefont {Opa\v{c}ak}, \citenamefont {Kainz}, \citenamefont {Sch\"{o}nhuber}, \citenamefont {Theiner}, \citenamefont {Limbacher}, \citenamefont {Beiser}, \citenamefont {Giparakis}, \citenamefont {Andrews}, \citenamefont {Strasser}, \citenamefont {Schwarz}, \citenamefont {Darmo},\ and\ \citenamefont {Unterrainer}}]{Jaidl21}%
  \BibitemOpen
  \bibfield  {author} {\bibinfo {author} {\bibfnamefont {M.}~\bibnamefont {Jaidl}}, \bibinfo {author} {\bibfnamefont {N.}~\bibnamefont {Opa\v{c}ak}}, \bibinfo {author} {\bibfnamefont {M.~A.}\ \bibnamefont {Kainz}}, \bibinfo {author} {\bibfnamefont {S.}~\bibnamefont {Sch\"{o}nhuber}}, \bibinfo {author} {\bibfnamefont {D.}~\bibnamefont {Theiner}}, \bibinfo {author} {\bibfnamefont {B.}~\bibnamefont {Limbacher}}, \bibinfo {author} {\bibfnamefont {M.}~\bibnamefont {Beiser}}, \bibinfo {author} {\bibfnamefont {M.}~\bibnamefont {Giparakis}}, \bibinfo {author} {\bibfnamefont {A.~M.}\ \bibnamefont {Andrews}}, \bibinfo {author} {\bibfnamefont {G.}~\bibnamefont {Strasser}}, \bibinfo {author} {\bibfnamefont {B.}~\bibnamefont {Schwarz}}, \bibinfo {author} {\bibfnamefont {J.}~\bibnamefont {Darmo}},\ and\ \bibinfo {author} {\bibfnamefont {K.}~\bibnamefont {Unterrainer}},\ }\bibfield  {title} {\bibinfo {title} {Comb operation in terahertz quantum cascade ring lasers},\ }\href {https://doi.org/10.1364/OPTICA.420674} {\bibfield
   {journal} {\bibinfo  {journal} {Optica}\ }\textbf {\bibinfo {volume} {8}},\ \bibinfo {pages} {780} (\bibinfo {year} {2021})}\BibitemShut {NoStop}%
\bibitem [{\citenamefont {Piccardo}\ \emph {et~al.}(2018{\natexlab{a}})\citenamefont {Piccardo}, \citenamefont {Chevalier}, \citenamefont {Mansuripur}, \citenamefont {Kazakov}, \citenamefont {Wang}, \citenamefont {Rubin}, \citenamefont {Meadowcroft}, \citenamefont {Belyanin},\ and\ \citenamefont {Capasso}}]{PiccardoHFCOptex}%
  \BibitemOpen
  \bibfield  {author} {\bibinfo {author} {\bibfnamefont {M.}~\bibnamefont {Piccardo}}, \bibinfo {author} {\bibfnamefont {P.}~\bibnamefont {Chevalier}}, \bibinfo {author} {\bibfnamefont {T.~S.}\ \bibnamefont {Mansuripur}}, \bibinfo {author} {\bibfnamefont {D.}~\bibnamefont {Kazakov}}, \bibinfo {author} {\bibfnamefont {Y.}~\bibnamefont {Wang}}, \bibinfo {author} {\bibfnamefont {N.~A.}\ \bibnamefont {Rubin}}, \bibinfo {author} {\bibfnamefont {L.}~\bibnamefont {Meadowcroft}}, \bibinfo {author} {\bibfnamefont {A.}~\bibnamefont {Belyanin}},\ and\ \bibinfo {author} {\bibfnamefont {F.}~\bibnamefont {Capasso}},\ }\bibfield  {title} {\bibinfo {title} {The harmonic state of quantum cascade lasers: origin, control, and prospective applications},\ }\href@noop {} {\bibfield  {journal} {\bibinfo  {journal} {Opt. Express}\ }\textbf {\bibinfo {volume} {26}},\ \bibinfo {pages} {9464} (\bibinfo {year} {2018}{\natexlab{a}})}\BibitemShut {NoStop}%
\bibitem [{\citenamefont {Wang}\ \emph {et~al.}(2020{\natexlab{b}})\citenamefont {Wang}, \citenamefont {Pistore}, \citenamefont {Riesch}, \citenamefont {Nong}, \citenamefont {Vigneron}, \citenamefont {Colombelli}, \citenamefont {Parillaud}, \citenamefont {Mangeney}, \citenamefont {Tignon}, \citenamefont {Jirauschek},\ and\ \citenamefont {Dhillon}}]{Dhillon1}%
  \BibitemOpen
  \bibfield  {author} {\bibinfo {author} {\bibfnamefont {F.}~\bibnamefont {Wang}}, \bibinfo {author} {\bibfnamefont {V.}~\bibnamefont {Pistore}}, \bibinfo {author} {\bibfnamefont {M.}~\bibnamefont {Riesch}}, \bibinfo {author} {\bibfnamefont {H.}~\bibnamefont {Nong}}, \bibinfo {author} {\bibfnamefont {P.-B.}\ \bibnamefont {Vigneron}}, \bibinfo {author} {\bibfnamefont {R.}~\bibnamefont {Colombelli}}, \bibinfo {author} {\bibfnamefont {O.}~\bibnamefont {Parillaud}}, \bibinfo {author} {\bibfnamefont {J.}~\bibnamefont {Mangeney}}, \bibinfo {author} {\bibfnamefont {J.}~\bibnamefont {Tignon}}, \bibinfo {author} {\bibfnamefont {C.}~\bibnamefont {Jirauschek}},\ and\ \bibinfo {author} {\bibfnamefont {S.~S.}\ \bibnamefont {Dhillon}},\ }\bibfield  {title} {\bibinfo {title} {Ultrafast response of harmonic modelocked {TH}z lasers},\ }\href@noop {} {\bibfield  {journal} {\bibinfo  {journal} {Light: Science {\&} Applications}\ }\textbf {\bibinfo {volume} {9}},\ \bibinfo {pages} {51} (\bibinfo {year}
  {2020}{\natexlab{b}})}\BibitemShut {NoStop}%
\bibitem [{\citenamefont {Piccardo}\ \emph {et~al.}(2018{\natexlab{b}})\citenamefont {Piccardo}, \citenamefont {Chevalier}, \citenamefont {Anand}, \citenamefont {Wang}, \citenamefont {Kazakov}, \citenamefont {Mejia}, \citenamefont {Xie}, \citenamefont {Lascola}, \citenamefont {Belyanin},\ and\ \citenamefont {Capasso}}]{PiccardoOptical}%
  \BibitemOpen
  \bibfield  {author} {\bibinfo {author} {\bibfnamefont {M.}~\bibnamefont {Piccardo}}, \bibinfo {author} {\bibfnamefont {P.}~\bibnamefont {Chevalier}}, \bibinfo {author} {\bibfnamefont {S.}~\bibnamefont {Anand}}, \bibinfo {author} {\bibfnamefont {Y.}~\bibnamefont {Wang}}, \bibinfo {author} {\bibfnamefont {D.}~\bibnamefont {Kazakov}}, \bibinfo {author} {\bibfnamefont {E.~A.}\ \bibnamefont {Mejia}}, \bibinfo {author} {\bibfnamefont {F.}~\bibnamefont {Xie}}, \bibinfo {author} {\bibfnamefont {K.}~\bibnamefont {Lascola}}, \bibinfo {author} {\bibfnamefont {A.}~\bibnamefont {Belyanin}},\ and\ \bibinfo {author} {\bibfnamefont {F.}~\bibnamefont {Capasso}},\ }\bibfield  {title} {\bibinfo {title} {Widely tunable harmonic frequency comb in a quantum cascade laser},\ }\href@noop {} {\bibfield  {journal} {\bibinfo  {journal} {Applied Physics Letters}\ }\textbf {\bibinfo {volume} {113}},\ \bibinfo {pages} {031104} (\bibinfo {year} {2018}{\natexlab{b}})}\BibitemShut {NoStop}%
\bibitem [{\citenamefont {Silvestri}\ \emph {et~al.}(2023{\natexlab{b}})\citenamefont {Silvestri}, \citenamefont {Qi}, \citenamefont {Taimre},\ and\ \citenamefont {Rakić}}]{Silvestri23}%
  \BibitemOpen
  \bibfield  {author} {\bibinfo {author} {\bibfnamefont {C.}~\bibnamefont {Silvestri}}, \bibinfo {author} {\bibfnamefont {X.}~\bibnamefont {Qi}}, \bibinfo {author} {\bibfnamefont {T.}~\bibnamefont {Taimre}},\ and\ \bibinfo {author} {\bibfnamefont {A.~D.}\ \bibnamefont {Rakić}},\ }\bibfield  {title} {\bibinfo {title} {Frequency combs induced by optical feedback and harmonic order tunability in quantum cascade lasers},\ }\href@noop {} {\bibfield  {journal} {\bibinfo  {journal} {APL Photonics}\ }\textbf {\bibinfo {volume} {8}},\ \bibinfo {pages} {116102} (\bibinfo {year} {2023}{\natexlab{b}})}\BibitemShut {NoStop}%
\bibitem [{\citenamefont {Silvestri}\ \emph {et~al.}(2025)\citenamefont {Silvestri}, \citenamefont {Qi}, \citenamefont {Taimre},\ and\ \citenamefont {Raki\ifmmode~\acute{c}\else \'{c}\fi{}}}]{SilvestriPRA2025}%
  \BibitemOpen
  \bibfield  {author} {\bibinfo {author} {\bibfnamefont {C.}~\bibnamefont {Silvestri}}, \bibinfo {author} {\bibfnamefont {X.}~\bibnamefont {Qi}}, \bibinfo {author} {\bibfnamefont {T.}~\bibnamefont {Taimre}},\ and\ \bibinfo {author} {\bibfnamefont {A.~D.}\ \bibnamefont {Raki\ifmmode~\acute{c}\else \'{c}\fi{}}},\ }\bibfield  {title} {\bibinfo {title} {Shaping terahertz harmonic frequency combs with frequency-dependent external reflectors},\ }\href {https://doi.org/10.1103/PhysRevA.111.043522} {\bibfield  {journal} {\bibinfo  {journal} {Phys. Rev. A}\ }\textbf {\bibinfo {volume} {111}},\ \bibinfo {pages} {043522} (\bibinfo {year} {2025})}\BibitemShut {NoStop}%
\bibitem [{\citenamefont {Riccardi}\ \emph {et~al.}(2024)\citenamefont {Riccardi}, \citenamefont {Guerrero}, \citenamefont {Pistore}, \citenamefont {Seitner}, \citenamefont {Jirauschek}, \citenamefont {Li}, \citenamefont {Davies}, \citenamefont {Linfield},\ and\ \citenamefont {Vitiello}}]{Riccardi24}%
  \BibitemOpen
  \bibfield  {author} {\bibinfo {author} {\bibfnamefont {E.}~\bibnamefont {Riccardi}}, \bibinfo {author} {\bibfnamefont {M.~A.~J.}\ \bibnamefont {Guerrero}}, \bibinfo {author} {\bibfnamefont {V.}~\bibnamefont {Pistore}}, \bibinfo {author} {\bibfnamefont {L.}~\bibnamefont {Seitner}}, \bibinfo {author} {\bibfnamefont {C.}~\bibnamefont {Jirauschek}}, \bibinfo {author} {\bibfnamefont {L.}~\bibnamefont {Li}}, \bibinfo {author} {\bibfnamefont {A.~G.}\ \bibnamefont {Davies}}, \bibinfo {author} {\bibfnamefont {E.~H.}\ \bibnamefont {Linfield}},\ and\ \bibinfo {author} {\bibfnamefont {M.~S.}\ \bibnamefont {Vitiello}},\ }\bibfield  {title} {\bibinfo {title} {Sculpting harmonic comb states in terahertz quantum cascade lasers by controlled engineering},\ }\href {https://doi.org/10.1364/OPTICA.509929} {\bibfield  {journal} {\bibinfo  {journal} {Optica}\ }\textbf {\bibinfo {volume} {11}},\ \bibinfo {pages} {412} (\bibinfo {year} {2024})}\BibitemShut {NoStop}%
\bibitem [{\citenamefont {Kazakov}\ \emph {et~al.}(2021)\citenamefont {Kazakov}, \citenamefont {Opa\v{c}ak}, \citenamefont {Beiser}, \citenamefont {Belyanin}, \citenamefont {Schwarz}, \citenamefont {Piccardo},\ and\ \citenamefont {Capasso}}]{Kazakov2021}%
  \BibitemOpen
  \bibfield  {author} {\bibinfo {author} {\bibfnamefont {D.}~\bibnamefont {Kazakov}}, \bibinfo {author} {\bibfnamefont {N.}~\bibnamefont {Opa\v{c}ak}}, \bibinfo {author} {\bibfnamefont {M.}~\bibnamefont {Beiser}}, \bibinfo {author} {\bibfnamefont {A.}~\bibnamefont {Belyanin}}, \bibinfo {author} {\bibfnamefont {B.}~\bibnamefont {Schwarz}}, \bibinfo {author} {\bibfnamefont {M.}~\bibnamefont {Piccardo}},\ and\ \bibinfo {author} {\bibfnamefont {F.}~\bibnamefont {Capasso}},\ }\bibfield  {title} {\bibinfo {title} {Defect-engineered ring laser harmonic frequency combs},\ }\href {https://doi.org/10.1364/OPTICA.430896} {\bibfield  {journal} {\bibinfo  {journal} {Optica}\ }\textbf {\bibinfo {volume} {8}},\ \bibinfo {pages} {1277} (\bibinfo {year} {2021})}\BibitemShut {NoStop}%
\bibitem [{\citenamefont {Piccardo}\ and\ \citenamefont {Capasso}(2022)}]{PiccardoReview}%
  \BibitemOpen
  \bibfield  {author} {\bibinfo {author} {\bibfnamefont {M.}~\bibnamefont {Piccardo}}\ and\ \bibinfo {author} {\bibfnamefont {F.}~\bibnamefont {Capasso}},\ }\bibfield  {title} {\bibinfo {title} {Laser frequency combs with fast gain recovery: Physics and applications},\ }\href@noop {} {\bibfield  {journal} {\bibinfo  {journal} {Laser \& Photonics Reviews}\ }\textbf {\bibinfo {volume} {16}},\ \bibinfo {pages} {2100403} (\bibinfo {year} {2022})}\BibitemShut {NoStop}%
\bibitem [{\citenamefont {Gordon}\ \emph {et~al.}(2008)\citenamefont {Gordon}, \citenamefont {Wang}, \citenamefont {Diehl}, \citenamefont {K\"artner}, \citenamefont {Belyanin}, \citenamefont {Bour}, \citenamefont {Corzine}, \citenamefont {H\"ofler}, \citenamefont {Liu}, \citenamefont {Schneider}, \citenamefont {Maier}, \citenamefont {Troccoli}, \citenamefont {Faist},\ and\ \citenamefont {Capasso}}]{gordon}%
  \BibitemOpen
  \bibfield  {author} {\bibinfo {author} {\bibfnamefont {A.}~\bibnamefont {Gordon}}, \bibinfo {author} {\bibfnamefont {C.~Y.}\ \bibnamefont {Wang}}, \bibinfo {author} {\bibfnamefont {L.}~\bibnamefont {Diehl}}, \bibinfo {author} {\bibfnamefont {F.~X.}\ \bibnamefont {K\"artner}}, \bibinfo {author} {\bibfnamefont {A.}~\bibnamefont {Belyanin}}, \bibinfo {author} {\bibfnamefont {D.}~\bibnamefont {Bour}}, \bibinfo {author} {\bibfnamefont {S.}~\bibnamefont {Corzine}}, \bibinfo {author} {\bibfnamefont {G.}~\bibnamefont {H\"ofler}}, \bibinfo {author} {\bibfnamefont {H.~C.}\ \bibnamefont {Liu}}, \bibinfo {author} {\bibfnamefont {H.}~\bibnamefont {Schneider}}, \bibinfo {author} {\bibfnamefont {T.}~\bibnamefont {Maier}}, \bibinfo {author} {\bibfnamefont {M.}~\bibnamefont {Troccoli}}, \bibinfo {author} {\bibfnamefont {J.}~\bibnamefont {Faist}},\ and\ \bibinfo {author} {\bibfnamefont {F.}~\bibnamefont {Capasso}},\ }\bibfield  {title} {\bibinfo {title} {Multimode regimes in quantum cascade lasers: From coherent
  instabilities to spatial hole burning},\ }\href {https://doi.org/10.1103/PhysRevA.77.053804} {\bibfield  {journal} {\bibinfo  {journal} {Phys. Rev. A}\ }\textbf {\bibinfo {volume} {77}},\ \bibinfo {pages} {053804} (\bibinfo {year} {2008})}\BibitemShut {NoStop}%
\bibitem [{\citenamefont {Vukovi\'c}\ \emph {et~al.}(2016)\citenamefont {Vukovi\'c}, \citenamefont {Radovanovi\'c}, \citenamefont {Milanovi\'c},\ and\ \citenamefont {Boiko}}]{Boiko1}%
  \BibitemOpen
  \bibfield  {author} {\bibinfo {author} {\bibfnamefont {N.}~\bibnamefont {Vukovi\'c}}, \bibinfo {author} {\bibfnamefont {J.}~\bibnamefont {Radovanovi\'c}}, \bibinfo {author} {\bibfnamefont {V.}~\bibnamefont {Milanovi\'c}},\ and\ \bibinfo {author} {\bibfnamefont {D.~L.}\ \bibnamefont {Boiko}},\ }\bibfield  {title} {\bibinfo {title} {Analytical expression for \uppercase{R}isken-\uppercase{N}ummedal-\uppercase{G}raham-\uppercase{H}aken instability threshold in quantum cascade lasers},\ }\href {https://doi.org/10.1364/OE.24.026911} {\bibfield  {journal} {\bibinfo  {journal} {Opt. Express}\ }\textbf {\bibinfo {volume} {24}},\ \bibinfo {pages} {26911} (\bibinfo {year} {2016})}\BibitemShut {NoStop}%
\bibitem [{\citenamefont {Vukovi\'c}\ \emph {et~al.}(2017)\citenamefont {Vukovi\'c}, \citenamefont {Radovanovi\'c}, \citenamefont {Milanovi\'c},\ and\ \citenamefont {Boiko}}]{Boiko2}%
  \BibitemOpen
  \bibfield  {author} {\bibinfo {author} {\bibfnamefont {N.~N.}\ \bibnamefont {Vukovi\'c}}, \bibinfo {author} {\bibfnamefont {J.}~\bibnamefont {Radovanovi\'c}}, \bibinfo {author} {\bibfnamefont {V.}~\bibnamefont {Milanovi\'c}},\ and\ \bibinfo {author} {\bibfnamefont {D.~L.}\ \bibnamefont {Boiko}},\ }\bibfield  {title} {\bibinfo {title} {Low-threshold \uppercase{RNGH} instabilities in quantum cascade lasers},\ }\href {https://doi.org/10.1109/JSTQE.2017.2699139} {\bibfield  {journal} {\bibinfo  {journal} {IEEE Journal of Selected Topics in Quantum Electronics}\ }\textbf {\bibinfo {volume} {23}},\ \bibinfo {pages} {1} (\bibinfo {year} {2017})}\BibitemShut {NoStop}%
\bibitem [{\citenamefont {Wang}\ and\ \citenamefont {Belyanin}(2020)}]{Belyanin2020}%
  \BibitemOpen
  \bibfield  {author} {\bibinfo {author} {\bibfnamefont {Y.}~\bibnamefont {Wang}}\ and\ \bibinfo {author} {\bibfnamefont {A.}~\bibnamefont {Belyanin}},\ }\bibfield  {title} {\bibinfo {title} {Harmonic frequency combs in quantum cascade lasers: Time-domain and frequency-domain theory},\ }\href {https://doi.org/10.1103/PhysRevA.102.013519} {\bibfield  {journal} {\bibinfo  {journal} {Phys. Rev. A}\ }\textbf {\bibinfo {volume} {102}},\ \bibinfo {pages} {013519} (\bibinfo {year} {2020})}\BibitemShut {NoStop}%
\bibitem [{\citenamefont {Risken}\ and\ \citenamefont {Nummedal}(1968)}]{Risken}%
  \BibitemOpen
  \bibfield  {author} {\bibinfo {author} {\bibfnamefont {H.}~\bibnamefont {Risken}}\ and\ \bibinfo {author} {\bibfnamefont {K.}~\bibnamefont {Nummedal}},\ }\bibfield  {title} {\bibinfo {title} {Self‐pulsing in lasers},\ }\href {https://doi.org/10.1063/1.1655817} {\bibfield  {journal} {\bibinfo  {journal} {Journal of Applied Physics}\ }\textbf {\bibinfo {volume} {39}},\ \bibinfo {pages} {4662} (\bibinfo {year} {1968})}\BibitemShut {NoStop}%
\bibitem [{\citenamefont {Lugiato}\ \emph {et~al.}(2015)\citenamefont {Lugiato}, \citenamefont {Prati},\ and\ \citenamefont {Brambilla}}]{Nos}%
  \BibitemOpen
  \bibfield  {author} {\bibinfo {author} {\bibfnamefont {L.}~\bibnamefont {Lugiato}}, \bibinfo {author} {\bibfnamefont {F.}~\bibnamefont {Prati}},\ and\ \bibinfo {author} {\bibfnamefont {M.}~\bibnamefont {Brambilla}},\ }\href {https://doi.org/10.1017/CBO9781107477254} {\emph {\bibinfo {title} {Nonlinear Optical Systems}}}\ (\bibinfo  {publisher} {Cambridge University Press},\ \bibinfo {year} {2015})\BibitemShut {NoStop}%
\bibitem [{\citenamefont {Bassi}\ \emph {et~al.}(2021)\citenamefont {Bassi}, \citenamefont {Prati},\ and\ \citenamefont {Lugiato}}]{Bassi2021}%
  \BibitemOpen
  \bibfield  {author} {\bibinfo {author} {\bibfnamefont {A.}~\bibnamefont {Bassi}}, \bibinfo {author} {\bibfnamefont {F.}~\bibnamefont {Prati}},\ and\ \bibinfo {author} {\bibfnamefont {L.~A.}\ \bibnamefont {Lugiato}},\ }\bibfield  {title} {\bibinfo {title} {Optical instabilities in {Fabry-Perot} resonators},\ }\bibfield  {journal} {\bibinfo  {journal} {Physical Review A}\ }\textbf {\bibinfo {volume} {103}},\ \href {https://doi.org/10.1103/PhysRevA.103.053519} {10.1103/PhysRevA.103.053519} (\bibinfo {year} {2021})\BibitemShut {NoStop}%
\bibitem [{\citenamefont {Lugiato}\ and\ \citenamefont {Prati}(2023)}]{Lugiato2023}%
  \BibitemOpen
  \bibfield  {author} {\bibinfo {author} {\bibfnamefont {L.~A.}\ \bibnamefont {Lugiato}}\ and\ \bibinfo {author} {\bibfnamefont {F.}~\bibnamefont {Prati}},\ }\bibfield  {title} {\bibinfo {title} {{Fabry–Perot} cavities made easy},\ }\href {https://doi.org/10.1016/bs.po.2023.02.001} {\bibfield  {journal} {\bibinfo  {journal} {Progress in Optics}\ }\textbf {\bibinfo {volume} {68}},\ \bibinfo {pages} {329 – 380} (\bibinfo {year} {2023})}\BibitemShut {NoStop}%
\bibitem [{\citenamefont {Lugiato}\ and\ \citenamefont {Prati}(2019)}]{Lugiato2019}%
  \BibitemOpen
  \bibfield  {author} {\bibinfo {author} {\bibfnamefont {L.~A.}\ \bibnamefont {Lugiato}}\ and\ \bibinfo {author} {\bibfnamefont {F.}~\bibnamefont {Prati}},\ }\bibfield  {title} {\bibinfo {title} {Self-pulsing in {Fabry-Perot} lasers: An analytic scenario},\ }\href@noop {} {\bibfield  {journal} {\bibinfo  {journal} {Phys. Rev. Res.}\ }\textbf {\bibinfo {volume} {1}} (\bibinfo {year} {2019})}\BibitemShut {NoStop}%
\bibitem [{\citenamefont {Prati}\ and\ \citenamefont {Columbo}(2007)}]{Prati2007}%
  \BibitemOpen
  \bibfield  {author} {\bibinfo {author} {\bibfnamefont {F.}~\bibnamefont {Prati}}\ and\ \bibinfo {author} {\bibfnamefont {L.}~\bibnamefont {Columbo}},\ }\bibfield  {title} {\bibinfo {title} {Long-wavelength instability in broad-area semiconductor lasers},\ }\href@noop {} {\bibfield  {journal} {\bibinfo  {journal} {Phys. Rev. A}\ }\textbf {\bibinfo {volume} {75}} (\bibinfo {year} {2007})}\BibitemShut {NoStop}%
\bibitem [{\citenamefont {Aranson}\ and\ \citenamefont {Kramer}(2002)}]{Aranson}%
  \BibitemOpen
  \bibfield  {author} {\bibinfo {author} {\bibfnamefont {I.~S.}\ \bibnamefont {Aranson}}\ and\ \bibinfo {author} {\bibfnamefont {L.}~\bibnamefont {Kramer}},\ }\bibfield  {title} {\bibinfo {title} {The world of the complex {Ginzburg-Landau} equation},\ }\href {https://doi.org/10.1103/RevModPhys.74.99} {\bibfield  {journal} {\bibinfo  {journal} {Rev. Mod. Phys.}\ }\textbf {\bibinfo {volume} {74}},\ \bibinfo {pages} {99} (\bibinfo {year} {2002})}\BibitemShut {NoStop}%
\bibitem [{\citenamefont {Columbo}\ \emph {et~al.}(2018{\natexlab{b}})\citenamefont {Columbo}, \citenamefont {Bardella},\ and\ \citenamefont {Gioannini}}]{Columbo18Qdot}%
  \BibitemOpen
  \bibfield  {author} {\bibinfo {author} {\bibfnamefont {L.~L.}\ \bibnamefont {Columbo}}, \bibinfo {author} {\bibfnamefont {P.}~\bibnamefont {Bardella}},\ and\ \bibinfo {author} {\bibfnamefont {M.}~\bibnamefont {Gioannini}},\ }\bibfield  {title} {\bibinfo {title} {Self-pulsing in single section ring lasers based on quantum dot materials: theory and simulations},\ }\href {https://doi.org/10.1364/OE.26.019044} {\bibfield  {journal} {\bibinfo  {journal} {Opt. Express}\ }\textbf {\bibinfo {volume} {26}},\ \bibinfo {pages} {19044} (\bibinfo {year} {2018}{\natexlab{b}})}\BibitemShut {NoStop}%
\bibitem [{\citenamefont {Silvestri}\ \emph {et~al.}(2024)\citenamefont {Silvestri}, \citenamefont {Brambilla}, \citenamefont {Bardella},\ and\ \citenamefont {Columbo}}]{silvestri2024unified}%
  \BibitemOpen
  \bibfield  {author} {\bibinfo {author} {\bibfnamefont {C.}~\bibnamefont {Silvestri}}, \bibinfo {author} {\bibfnamefont {M.}~\bibnamefont {Brambilla}}, \bibinfo {author} {\bibfnamefont {P.}~\bibnamefont {Bardella}},\ and\ \bibinfo {author} {\bibfnamefont {L.~L.}\ \bibnamefont {Columbo}},\ }\bibfield  {title} {\bibinfo {title} {{Unified theory for frequency combs in ring and {Fabry–Perot} quantum cascade lasers: An order-parameter equation approach}},\ }\href@noop {} {\bibfield  {journal} {\bibinfo  {journal} {APL Photonics}\ }\textbf {\bibinfo {volume} {9}},\ \bibinfo {pages} {076119} (\bibinfo {year} {2024})}\BibitemShut {NoStop}%
\bibitem [{\citenamefont {Torrey}(1949)}]{Torrey49}%
  \BibitemOpen
  \bibfield  {author} {\bibinfo {author} {\bibfnamefont {H.~C.}\ \bibnamefont {Torrey}},\ }\bibfield  {title} {\bibinfo {title} {Transient nutations in nuclear magnetic resonance},\ }\href {https://doi.org/10.1103/PhysRev.76.1059} {\bibfield  {journal} {\bibinfo  {journal} {Phys. Rev.}\ }\textbf {\bibinfo {volume} {76}},\ \bibinfo {pages} {1059} (\bibinfo {year} {1949})}\BibitemShut {NoStop}%
\bibitem [{\citenamefont {Jumpertz}\ \emph {et~al.}(2016)\citenamefont {Jumpertz}, \citenamefont {Michel}, \citenamefont {Pawlus}, \citenamefont {Elsässer}, \citenamefont {Schires}, \citenamefont {Carras},\ and\ \citenamefont {Grillot}}]{Grillot16}%
  \BibitemOpen
  \bibfield  {author} {\bibinfo {author} {\bibfnamefont {L.}~\bibnamefont {Jumpertz}}, \bibinfo {author} {\bibfnamefont {F.}~\bibnamefont {Michel}}, \bibinfo {author} {\bibfnamefont {R.}~\bibnamefont {Pawlus}}, \bibinfo {author} {\bibfnamefont {W.}~\bibnamefont {Elsässer}}, \bibinfo {author} {\bibfnamefont {K.}~\bibnamefont {Schires}}, \bibinfo {author} {\bibfnamefont {M.}~\bibnamefont {Carras}},\ and\ \bibinfo {author} {\bibfnamefont {F.}~\bibnamefont {Grillot}},\ }\bibfield  {title} {\bibinfo {title} {Measurements of the linewidth enhancement factor of mid-infrared quantum cascade lasers by different optical feedback techniques},\ }\href@noop {} {\bibfield  {journal} {\bibinfo  {journal} {AIP Advances}\ }\textbf {\bibinfo {volume} {6}},\ \bibinfo {pages} {015212} (\bibinfo {year} {2016})}\BibitemShut {NoStop}%
\bibitem [{\citenamefont {Capua}\ \emph {et~al.}(2014)\citenamefont {Capua}, \citenamefont {Karni}, \citenamefont {Eisenstein}, \citenamefont {Sichkovskyi}, \citenamefont {Ivanov},\ and\ \citenamefont {Reithmaier}}]{Capua2014}%
  \BibitemOpen
  \bibfield  {author} {\bibinfo {author} {\bibfnamefont {A.}~\bibnamefont {Capua}}, \bibinfo {author} {\bibfnamefont {O.}~\bibnamefont {Karni}}, \bibinfo {author} {\bibfnamefont {G.}~\bibnamefont {Eisenstein}}, \bibinfo {author} {\bibfnamefont {V.}~\bibnamefont {Sichkovskyi}}, \bibinfo {author} {\bibfnamefont {V.}~\bibnamefont {Ivanov}},\ and\ \bibinfo {author} {\bibfnamefont {J.~P.}\ \bibnamefont {Reithmaier}},\ }\bibfield  {title} {\bibinfo {title} {Coherent control in a semiconductor optical amplifier operating at room temperature},\ }\href {https://doi.org/10.1038/ncomms6025} {\bibfield  {journal} {\bibinfo  {journal} {Nature Communications}\ }\textbf {\bibinfo {volume} {5}},\ \bibinfo {pages} {5025} (\bibinfo {year} {2014})}\BibitemShut {NoStop}%
\bibitem [{\citenamefont {Liu}\ \emph {et~al.}(2010)\citenamefont {Liu}, \citenamefont {Kumar}, \citenamefont {Huybrechts}, \citenamefont {Spuesens}, \citenamefont {Roelkens}, \citenamefont {Geluk}, \citenamefont {de~Vries}, \citenamefont {Regreny}, \citenamefont {Van~Thourhout}, \citenamefont {Baets},\ and\ \citenamefont {Morthier}}]{Liu2010}%
  \BibitemOpen
  \bibfield  {author} {\bibinfo {author} {\bibfnamefont {L.}~\bibnamefont {Liu}}, \bibinfo {author} {\bibfnamefont {R.}~\bibnamefont {Kumar}}, \bibinfo {author} {\bibfnamefont {K.}~\bibnamefont {Huybrechts}}, \bibinfo {author} {\bibfnamefont {T.}~\bibnamefont {Spuesens}}, \bibinfo {author} {\bibfnamefont {G.}~\bibnamefont {Roelkens}}, \bibinfo {author} {\bibfnamefont {E.-J.}\ \bibnamefont {Geluk}}, \bibinfo {author} {\bibfnamefont {T.}~\bibnamefont {de~Vries}}, \bibinfo {author} {\bibfnamefont {P.}~\bibnamefont {Regreny}}, \bibinfo {author} {\bibfnamefont {D.}~\bibnamefont {Van~Thourhout}}, \bibinfo {author} {\bibfnamefont {R.}~\bibnamefont {Baets}},\ and\ \bibinfo {author} {\bibfnamefont {G.}~\bibnamefont {Morthier}},\ }\bibfield  {title} {\bibinfo {title} {An ultra-small, low-power, all-optical flip-flop memory on a silicon chip},\ }\href {https://doi.org/10.1038/nphoton.2009.268} {\bibfield  {journal} {\bibinfo  {journal} {Nature Photonics}\ }\textbf {\bibinfo {volume} {4}},\ \bibinfo {pages} {182}
  (\bibinfo {year} {2010})}\BibitemShut {NoStop}%
\bibitem [{\citenamefont {Choi}\ \emph {et~al.}(2010)\citenamefont {Choi}, \citenamefont {Gkortsas}, \citenamefont {Diehl}, \citenamefont {Bour}, \citenamefont {Corzine}, \citenamefont {Zhu}, \citenamefont {H{\"o}fler}, \citenamefont {Capasso}, \citenamefont {K{\"a}rtner},\ and\ \citenamefont {Norris}}]{Choi2010}%
  \BibitemOpen
  \bibfield  {author} {\bibinfo {author} {\bibfnamefont {H.}~\bibnamefont {Choi}}, \bibinfo {author} {\bibfnamefont {V.-M.}\ \bibnamefont {Gkortsas}}, \bibinfo {author} {\bibfnamefont {L.}~\bibnamefont {Diehl}}, \bibinfo {author} {\bibfnamefont {D.}~\bibnamefont {Bour}}, \bibinfo {author} {\bibfnamefont {S.}~\bibnamefont {Corzine}}, \bibinfo {author} {\bibfnamefont {J.}~\bibnamefont {Zhu}}, \bibinfo {author} {\bibfnamefont {G.}~\bibnamefont {H{\"o}fler}}, \bibinfo {author} {\bibfnamefont {F.}~\bibnamefont {Capasso}}, \bibinfo {author} {\bibfnamefont {F.~X.}\ \bibnamefont {K{\"a}rtner}},\ and\ \bibinfo {author} {\bibfnamefont {T.~B.}\ \bibnamefont {Norris}},\ }\bibfield  {title} {\bibinfo {title} {Ultrafast rabi flopping and coherent pulse propagation in a quantum cascade laser},\ }\href {https://doi.org/10.1038/nphoton.2010.205} {\bibfield  {journal} {\bibinfo  {journal} {Nature Photonics}\ }\textbf {\bibinfo {volume} {4}},\ \bibinfo {pages} {706} (\bibinfo {year} {2010})}\BibitemShut {NoStop}%
\bibitem [{\citenamefont {Sumetsky}\ \emph {et~al.}(2010)\citenamefont {Sumetsky}, \citenamefont {Dulashko},\ and\ \citenamefont {Windeler}}]{Sumetsky10}%
  \BibitemOpen
  \bibfield  {author} {\bibinfo {author} {\bibfnamefont {M.}~\bibnamefont {Sumetsky}}, \bibinfo {author} {\bibfnamefont {Y.}~\bibnamefont {Dulashko}},\ and\ \bibinfo {author} {\bibfnamefont {R.~S.}\ \bibnamefont {Windeler}},\ }\bibfield  {title} {\bibinfo {title} {Super free spectral range tunable optical microbubble resonator},\ }\href {https://doi.org/10.1364/OL.35.001866} {\bibfield  {journal} {\bibinfo  {journal} {Opt. Lett.}\ }\textbf {\bibinfo {volume} {35}},\ \bibinfo {pages} {1866} (\bibinfo {year} {2010})}\BibitemShut {NoStop}%
\bibitem [{\citenamefont {Krasnokutska}\ \emph {et~al.}(2019)\citenamefont {Krasnokutska}, \citenamefont {Tambasco},\ and\ \citenamefont {Peruzzo}}]{Krasnokutska2019}%
  \BibitemOpen
  \bibfield  {author} {\bibinfo {author} {\bibfnamefont {I.}~\bibnamefont {Krasnokutska}}, \bibinfo {author} {\bibfnamefont {J.-L.~J.}\ \bibnamefont {Tambasco}},\ and\ \bibinfo {author} {\bibfnamefont {A.}~\bibnamefont {Peruzzo}},\ }\bibfield  {title} {\bibinfo {title} {Tunable large free spectral range microring resonators in lithium niobate on insulator},\ }\href {https://doi.org/10.1038/s41598-019-47231-3} {\bibfield  {journal} {\bibinfo  {journal} {Scientific Reports}\ }\textbf {\bibinfo {volume} {9}},\ \bibinfo {pages} {11086} (\bibinfo {year} {2019})}\BibitemShut {NoStop}%
\end{thebibliography}%


\begin{thebibliography}{6}%
\makeatletter
\providecommand \@ifxundefined [1]{%
 \@ifx{#1\undefined}
}%
\providecommand \@ifnum [1]{%
 \ifnum #1\expandafter \@firstoftwo
 \else \expandafter \@secondoftwo
 \fi
}%
\providecommand \@ifx [1]{%
 \ifx #1\expandafter \@firstoftwo
 \else \expandafter \@secondoftwo
 \fi
}%
\providecommand \natexlab [1]{#1}%
\providecommand \enquote  [1]{``#1''}%
\providecommand \bibnamefont  [1]{#1}%
\providecommand \bibfnamefont [1]{#1}%
\providecommand \citenamefont [1]{#1}%
\providecommand \href@noop [0]{\@secondoftwo}%
\providecommand \href [0]{\begingroup \@sanitize@url \@href}%
\providecommand \@href[1]{\@@startlink{#1}\@@href}%
\providecommand \@@href[1]{\endgroup#1\@@endlink}%
\providecommand \@sanitize@url [0]{\catcode `\\12\catcode `\$12\catcode `\&12\catcode `\#12\catcode `\^12\catcode `\_12\catcode `\%12\relax}%
\providecommand \@@startlink[1]{}%
\providecommand \@@endlink[0]{}%
\providecommand \url  [0]{\begingroup\@sanitize@url \@url }%
\providecommand \@url [1]{\endgroup\@href {#1}{\urlprefix }}%
\providecommand \urlprefix  [0]{URL }%
\providecommand \Eprint [0]{\href }%
\providecommand \doibase [0]{https://doi.org/}%
\providecommand \selectlanguage [0]{\@gobble}%
\providecommand \bibinfo  [0]{\@secondoftwo}%
\providecommand \bibfield  [0]{\@secondoftwo}%
\providecommand \translation [1]{[#1]}%
\providecommand \BibitemOpen [0]{}%
\providecommand \bibitemStop [0]{}%
\providecommand \bibitemNoStop [0]{.\EOS\space}%
\providecommand \EOS [0]{\spacefactor3000\relax}%
\providecommand \BibitemShut  [1]{\csname bibitem#1\endcsname}%
\let\auto@bib@innerbib\@empty
\bibitem [{\citenamefont {Columbo}\ \emph {et~al.}(2018{\natexlab{a}})\citenamefont {Columbo}, \citenamefont {Barbieri}, \citenamefont {Sirtori},\ and\ \citenamefont {Brambilla}}]{Columbo2018}%
  \BibitemOpen
  \bibfield  {author} {\bibinfo {author} {\bibfnamefont {L.~L.}\ \bibnamefont {Columbo}}, \bibinfo {author} {\bibfnamefont {S.}~\bibnamefont {Barbieri}}, \bibinfo {author} {\bibfnamefont {C.}~\bibnamefont {Sirtori}},\ and\ \bibinfo {author} {\bibfnamefont {M.}~\bibnamefont {Brambilla}},\ }\bibfield  {title} {\bibinfo {title} {Dynamics of a broad-band quantum cascade laser: from chaos to coherent dynamics and mode-locking},\ }\href {http://opg.optica.org/oe/abstract.cfm?URI=oe-26-3-2829} {\bibfield  {journal} {\bibinfo  {journal} {Opt. Express}\ }\textbf {\bibinfo {volume} {26}},\ \bibinfo {pages} {2829} (\bibinfo {year} {2018}{\natexlab{a}})}\BibitemShut {NoStop}%
\bibitem [{\citenamefont {Lugiato}\ \emph {et~al.}(2015)\citenamefont {Lugiato}, \citenamefont {Prati},\ and\ \citenamefont {Brambilla}}]{NOS}%
  \BibitemOpen
  \bibfield  {author} {\bibinfo {author} {\bibfnamefont {L.}~\bibnamefont {Lugiato}}, \bibinfo {author} {\bibfnamefont {F.}~\bibnamefont {Prati}},\ and\ \bibinfo {author} {\bibfnamefont {M.}~\bibnamefont {Brambilla}},\ }\href {https://doi.org/10.1017/CBO9781107477254} {\emph {\bibinfo {title} {Nonlinear Optical Systems}}}\ (\bibinfo  {publisher} {Cambridge University Press},\ \bibinfo {year} {2015})\BibitemShut {NoStop}%
\bibitem [{\citenamefont {Columbo}\ \emph {et~al.}(2018{\natexlab{b}})\citenamefont {Columbo}, \citenamefont {Bardella},\ and\ \citenamefont {Gioannini}}]{Columbo18Qdot}%
  \BibitemOpen
  \bibfield  {author} {\bibinfo {author} {\bibfnamefont {L.~L.}\ \bibnamefont {Columbo}}, \bibinfo {author} {\bibfnamefont {P.}~\bibnamefont {Bardella}},\ and\ \bibinfo {author} {\bibfnamefont {M.}~\bibnamefont {Gioannini}},\ }\bibfield  {title} {\bibinfo {title} {Self-pulsing in single section ring lasers based on quantum dot materials: theory and simulations},\ }\href {https://doi.org/10.1364/OE.26.019044} {\bibfield  {journal} {\bibinfo  {journal} {Opt. Express}\ }\textbf {\bibinfo {volume} {26}},\ \bibinfo {pages} {19044} (\bibinfo {year} {2018}{\natexlab{b}})}\BibitemShut {NoStop}%
\bibitem [{\citenamefont {Torrey}(1949)}]{Torrey49}%
  \BibitemOpen
  \bibfield  {author} {\bibinfo {author} {\bibfnamefont {H.~C.}\ \bibnamefont {Torrey}},\ }\bibfield  {title} {\bibinfo {title} {Transient nutations in nuclear magnetic resonance},\ }\href {https://doi.org/10.1103/PhysRev.76.1059} {\bibfield  {journal} {\bibinfo  {journal} {Phys. Rev.}\ }\textbf {\bibinfo {volume} {76}},\ \bibinfo {pages} {1059} (\bibinfo {year} {1949})}\BibitemShut {NoStop}%
\bibitem [{\citenamefont {Aranson}\ and\ \citenamefont {Kramer}(2002)}]{Aranson}%
  \BibitemOpen
  \bibfield  {author} {\bibinfo {author} {\bibfnamefont {I.~S.}\ \bibnamefont {Aranson}}\ and\ \bibinfo {author} {\bibfnamefont {L.}~\bibnamefont {Kramer}},\ }\bibfield  {title} {\bibinfo {title} {The world of the complex {Ginzburg-Landau} equation},\ }\href {https://doi.org/10.1103/RevModPhys.74.99} {\bibfield  {journal} {\bibinfo  {journal} {Rev. Mod. Phys.}\ }\textbf {\bibinfo {volume} {74}},\ \bibinfo {pages} {99} (\bibinfo {year} {2002})}\BibitemShut {NoStop}%
\bibitem [{\citenamefont {Piccardo}\ \emph {et~al.}(2020)\citenamefont {Piccardo}, \citenamefont {Schwarz}, \citenamefont {Kazakov}, \citenamefont {Beiser}, \citenamefont {Opa{\v{c}}ak}, \citenamefont {Wang}, \citenamefont {Jha}, \citenamefont {Hillbrand}, \citenamefont {Tamagnone}, \citenamefont {Chen}, \citenamefont {Zhu}, \citenamefont {Columbo}, \citenamefont {Belyanin},\ and\ \citenamefont {Capasso}}]{NaturePiccardo}%
  \BibitemOpen
  \bibfield  {author} {\bibinfo {author} {\bibfnamefont {M.}~\bibnamefont {Piccardo}}, \bibinfo {author} {\bibfnamefont {B.}~\bibnamefont {Schwarz}}, \bibinfo {author} {\bibfnamefont {D.}~\bibnamefont {Kazakov}}, \bibinfo {author} {\bibfnamefont {M.}~\bibnamefont {Beiser}}, \bibinfo {author} {\bibfnamefont {N.}~\bibnamefont {Opa{\v{c}}ak}}, \bibinfo {author} {\bibfnamefont {Y.}~\bibnamefont {Wang}}, \bibinfo {author} {\bibfnamefont {S.}~\bibnamefont {Jha}}, \bibinfo {author} {\bibfnamefont {J.}~\bibnamefont {Hillbrand}}, \bibinfo {author} {\bibfnamefont {M.}~\bibnamefont {Tamagnone}}, \bibinfo {author} {\bibfnamefont {W.~T.}\ \bibnamefont {Chen}}, \bibinfo {author} {\bibfnamefont {A.~Y.}\ \bibnamefont {Zhu}}, \bibinfo {author} {\bibfnamefont {L.~L.}\ \bibnamefont {Columbo}}, \bibinfo {author} {\bibfnamefont {A.}~\bibnamefont {Belyanin}},\ and\ \bibinfo {author} {\bibfnamefont {F.}~\bibnamefont {Capasso}},\ }\bibfield  {title} {\bibinfo {title} {Frequency combs induced by phase turbulence},\ }\href
  {https://doi.org/10.1038/s41586-020-2386-6} {\bibfield  {journal} {\bibinfo  {journal} {Nature}\ }\textbf {\bibinfo {volume} {582}},\ \bibinfo {pages} {360} (\bibinfo {year} {2020})}\BibitemShut {NoStop}%
\end{thebibliography}%
\end{document}


\preprint{APS/123-QED}

\title{Effective Rabi frequency in semiconductor lasers and the origin of self-starting harmonic frequency combs -- Supplementary materials}%

\author{Carlo Silvestri}
\affiliation{Institute of Photonics and Optical Science (IPOS), School of Physics, The University of Sydney, NSW 2006, Australia
}
\author{Franco Prati}%
\affiliation{Dipartimento di Scienza e Alta Tecnologia,
Universit\`a dell’Insubria, 22100 Como, Italy
}
\author{Massimo Brambilla}
\affiliation{ Dipartimento Interateneo di Fisica, Politecnico di Bari and CNR-IFN, UOS Bari, Italy
}
\author{Mariangela Gioannini}
\affiliation{Dipartimento di Elettronica e Telecomunicazioni, Politecnico di Torino, 10129 Torino, Italy
}
\author{Lorenzo Luigi Columbo}
\affiliation{Dipartimento di Elettronica e Telecomunicazioni, Politecnico di Torino, 10129 Torino, Italy
}
\maketitle
\section{Continuous wave solutions of the Effective semiconductor Maxwell-Bloch equations}
Let us consider the  ESMBEs for the ring configuration:
\begin{eqnarray}
\frac{\partial F}{\partial \eta} + \frac{\partial F}{\partial t} &=& -\sigma\left(F+P\right)\,,\label{eq:F}\\
\frac{\partial P}{\partial t} &=& -\Gamma (1 + i\alpha)^2FD-\Gamma (1 + i\alpha)P\,,\label{eq:P}\\
\frac{\partial D}{\partial t} &=& b \left[\mu - D + \frac{1}{2}\left(F^* P + F P^*\right)\right]\,.\label{eq:D}
\end{eqnarray}
Following the approach presented in \cite{Columbo2018}, we seek solutions of the form $F = F_0 e^{-ik\eta + i\omega t}$, $P = P_0 e^{-ik\eta + i\omega t}$, $D = D_0$. We then obtain:
\begin{eqnarray}
-ikF_0 + i\omega F_0 &=& -\sigma (F_0 + P_0)\,,\label{eq:F0}\\
i\omega P_0 &=& -\Gamma (1 + i\alpha)^2F_0D_0-\Gamma (1 + i\alpha) P_0\,,\label{eq:P0}\\
0 &=& \mu - D_0 + \frac{1}{2}\left(F_0^* P_0 + F_0 P_0^*\right)\,.\label{eq:D0}
\end{eqnarray}
From Eq.~\eqref{eq:P0} we can write an expression for $P_0$:
\begin{equation}\label{eq:P0_3}
P_0 = -\frac{\Gamma (1 -\alpha^2) + 2i\alpha\Gamma}{\Gamma + i(\omega + \alpha \Gamma)} D_0 F_0=\left[-H_1(\omega)+iH_2(\omega)\right]D_0 F_0\,,
\end{equation}
with
\begin{eqnarray}
H_1 (\omega) &=& \frac{\Gamma^2 (1 - \alpha^2) + 2\alpha \Gamma (\omega + \alpha \Gamma)}{\Gamma^2 + (\omega + \alpha \Gamma)^2}\,,\label{eq:H1}\\
H_2 (\omega) &=& \frac{-2\alpha \Gamma^2 + \Gamma(1 - \alpha^2) (\omega + \alpha \Gamma)}{\Gamma^2 + (\omega + \alpha \Gamma)^2}\,.\label{eq:H2}
\end{eqnarray}
Substituting Eq.~\eqref{eq:P0_3} in Eqs.~\eqref{eq:F0} and \eqref{eq:D0} we obtain:
\begin{eqnarray}
F_0 \left(ik - i\omega - \sigma\right) &=& \sigma\left[-H_1 (\omega) + iH_2 (\omega)\right]D_0 F_0\,,\label{eq:F0_3}\\
0 &=& \mu -D_0 -H_1(\omega)|F_0|^2 D_0\,.\label{eq:D0_2}
\end{eqnarray}
Dividing Eq.~\eqref{eq:F0_3} by $F_0$ and splitting it into real and imaginary parts, we obtain from the real part:
\begin{equation}\label{eq:D0_5}
D_0 = \frac{1}{H_1 (\omega)}\,,
\end{equation}
and from the imaginary part:
\begin{equation}\label{eq:omega_ss}
\omega = k - \sigma\frac{H_2 (\omega)}{H_1 (\omega)}\,.
\end{equation}
Eq.~\eqref{eq:D0_2} gives an expression for $D_0$
\begin{equation}
D_0=\frac{\mu}{1+|F_0|^2H_1(\omega)}\,,
\end{equation}
which, combined with Eq.~\eqref{eq:D0_5}, allows to write the intensity as
\begin{equation}\label{eq:F0_squared_ss}
|F_0|^2=\mu-\frac{1}{H_1(\omega)}\,.
\end{equation}
We highlight that Eq.~\eqref{eq:F0_squared_ss} describes the dependence of the field intensity on the pump parameter, while Eq.~\eqref{eq:omega_ss} represents the dispersion relation, which, as can be noticed, is a nonlinear equation.
\section{Linear stability analysis}
In this section we perform the linear stability analysis of the CW solutions Eqs.~\eqref{eq:P0_3}, \eqref{eq:D0_5}, \eqref{eq:F0_squared_ss}, introducing time dependent perturbations for the dynamical variables $F$, $P$, and $D$ \cite{NOS,Columbo18Qdot}:
\begin{align}
    F &= (F_0 + \delta F) e^{-ik\eta + i\omega t}, \label{eq:F_expansion} \\
    P &= (P_0 + \delta P) e^{-ik\eta + i\omega t}, \label{eq:P_expansion} \\
    D &= D_0 + \delta D. \label{eq:D_expansion}
\end{align}
We substitute Eqs.~\eqref{eq:F_expansion}--\eqref{eq:D_expansion} into Eqs.~\eqref{eq:F}--\eqref{eq:D}, drop the nonlinear terms 
and expand the generic dynamical variable perturbation $\delta X$ as:
\begin{equation}
    \delta X = \sum_n \delta X_n e^{-ik_n \eta}e^{\lambda_n t}, \label{eq:deltaF_expansion}
\end{equation}
where $k_\mathrm{n}$ represent the wavenumbers of the mode expansion for each of the dynamical variables. 
Then, for each mode index $n$, we obtain a closed subset of equations
\begin{eqnarray}
    \delta F_n [\sigma-i(k - \omega + k_n) + \lambda_n] + \sigma \delta P_n &=& 0\,, \label{eq:deltaF_n_eq}\\
    \delta F_{-n}^* [\sigma+i(k - \omega - k_n) + \lambda_n ] + \sigma \delta P_{-n}^* &=& 0\,,\label{eq:deltaF_minus_n_eq}\\
    \delta P_n [i\omega + \Gamma(1 + i\alpha) + \lambda_n] + \Gamma(1 + i\alpha)^2(D_0 \delta F_n + F_0 \delta D_n) &=& 0\,, \label{eq:deltaP_n_eq}\\
    \delta P_{-n}^* [-i\omega + \Gamma(1 - i\alpha) + \lambda_n] + \Gamma(1 - i\alpha)^2 (D_0 \delta F_{-n}^* + F_0^* \delta D_n) &=& 0\,, \label{eq:deltaP_minus_n_eq}\\
    \delta D_n (\lambda_n + b) - \frac{b}{2}\left( P_0 \delta F_{-n}^* + F_0^* \delta P_n + F_0 \delta P_{-n}^* + P_0^* \delta F_n\right) &=& 0\,. \label{eq:deltaD_n_eq}
\end{eqnarray}

From Eqs.~\eqref{eq:deltaF_n_eq}--\eqref{eq:deltaD_n_eq}, we can construct the matrix $M_{\lambda_n}$ of the linearized system:

\begin{equation}
M_{\lambda_n} =
\begin{pmatrix}
\sigma-i(k - \omega + k_n)  + \lambda_n & 0 & \sigma & 0 & 0 \\
0 & \sigma+i(k - \omega - k_n) + \lambda_n & 0 & \sigma & 0 \\
\Gamma(1 + i\alpha)^2D_0 & 0 & i\omega + \Gamma(1 + i\alpha) + \lambda_n & 0 & \Gamma(1 + i\alpha)^2F_0 \\
0 & \Gamma(1 - i\alpha)^2D_0 & 0& -i\omega + \Gamma(1 - i\alpha) + \lambda_n & \Gamma(1 - i\alpha)^2F_0^* \\
-\frac{b}{2} P_0^* & -\frac{b}{2}P_0 & -\frac{b}{2}F_0^* & -\frac{b}{2}F_0 & b + \lambda_n
\end{pmatrix}\,.
\label{Secmatrix}
\end{equation}
The eigenvalues $\lambda_n$ are determined by solving the characteristic equation $\det(M_{\lambda_n}) = 0$.
\section{Effective Rabi Frequency}
We assume that the field $F$ is constant and monochromatic with a frequency coincident with the maximum of the unsaturated gain
and consider the dynamical equations for $P$, $P^*$, and $D$
\begin{align}
\frac{dP}{dt} &= -\Gamma(1 + i\alpha)^{2}DF-\Gamma(1 + i\alpha)P, \label{Peq_a} \\
\frac{dP^*}{dt} &= -\Gamma(1 - i\alpha)^{2}DF-\Gamma(1 - i\alpha)P^*, \label{Peq_b} \\
\frac{dD}{dt} &= b\mu -b D + \frac{b}{2}\left(F^* P + F P^*\right). \label{Deq_a}
\end{align}
The dynamics of this linear system is described by the complex eigenvalues $\beta$ solutions of the characteristic equation 
$\det(M_\beta) = 0$ with
\begin{equation}\label{eqcarattbeta}
M_\beta=
\begin{pmatrix}
\beta + \Gamma(1+i\alpha) & 0 & \Gamma (1+i\alpha)^2F\\
0& \beta + \Gamma(1-i\alpha) & \Gamma (1-i\alpha)^2F^* \\
-\frac{b}{2}F^*&  -\frac{b}{2}F &  \beta+b
\end{pmatrix}\,.
\end{equation}
The cubic characteristic equation is
\begin{align}\label{beta_caratt}
\beta^3+(2 \Gamma + b) \beta^2+ \Gamma\left[ \Gamma (1 + \alpha^2) + b (2 + X - X \alpha^2) \right] \beta +b\Gamma^2 (1 + \alpha^2)(1+X) =0\,, 
\end{align}
where we set $X=|F|^2$. As mentioned in the main text, Eq.~\eqref{beta_caratt} has in general two complex-conjugate solutions and one real solution. Therefore, we identify the Effective Rabi frequency (ERF) with the absolute value of the imaginary part of the complex solutions. In physical units, the expression of the Rabi frequency is ERF$=|Im(\beta)|/(2\pi\tau_d)$.
If $\alpha=0$ (two-level atoms) the characteristic equation can be factorized as
\begin{align}
(\beta+\Gamma)\left[\beta^2+(\Gamma+b)\beta+b\Gamma(1+X)\right]=0\,,
\end{align}
whose solutions are
\begin{align}
    \beta_1=-\Gamma\,,\qquad \beta_\pm=\frac{1}{2}\left[-\Gamma-b\pm\sqrt{(\Gamma-b)^2-4b\Gamma X}\right]\,.
\end{align}
The latter are complex if $X>(\Gamma-b)^2/(4b\Gamma)$ and the associated ERF is
\begin{align}
    \mathrm{Im}(\beta_+)=\sqrt{4b\Gamma X-(\Gamma-b)^2}\,,
\end{align}
as for two-level lasers in resonance \cite{Torrey49}.
At the laser threshold ($X=0$) the characteristic equation can be factorized as well
\begin{align}
(\beta+b)\left[\Gamma^2\alpha^2+(\Gamma+\beta)^2\right]=0\,.
\end{align}
The solutions are
\begin{align}
    \beta_1=-b\,,\qquad \beta_\pm=-\Gamma\pm i\Gamma\alpha\,.
\end{align}
and the associated ERF is
\begin{align}
    \mathrm{Im}(\beta_+)=\Gamma\alpha\,.
\end{align}
This explains the behaviour reported in Fig. 1(a) of the main text, where the ERF at $X=0$ increases proportionally to $\alpha$.

\newpage
\section{Scan of the cavity length for $\alpha>1$}
\begin{figure}[h!]
\begin{center}
    \includegraphics[width=0.75\textwidth]{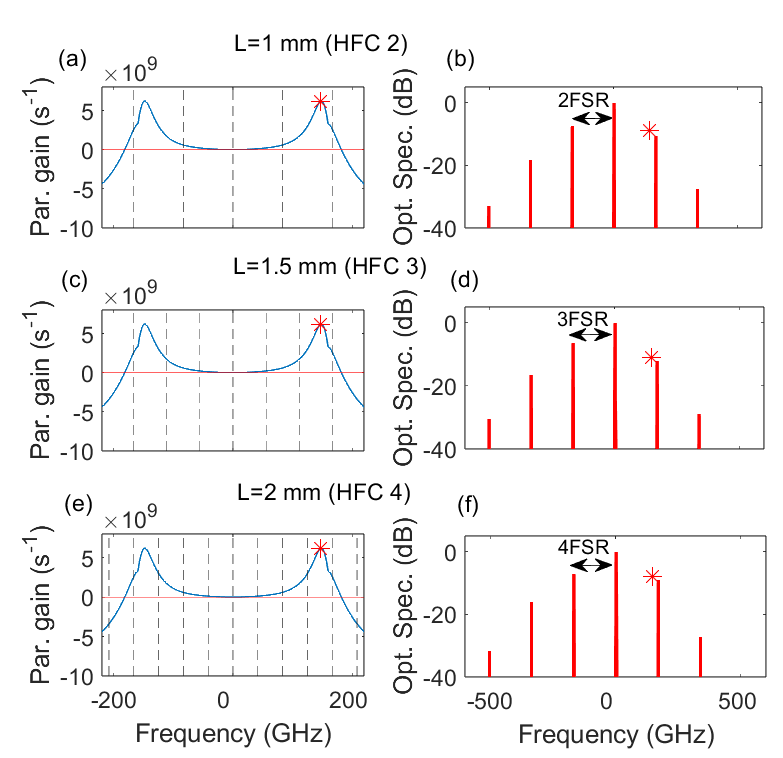}
       \end{center}

    \caption{Harmonic frequency combs of different order obtained by scanning the cavity length L at fixed $\alpha=1.05$. The other parameters are $\Gamma=0.06$, $\mu=7.9$, $\sigma=1.6\times10^{-3}$, and $b=0.02$.  Parametric gain as a function of the frequency (panels (a), (c), (e)), and simulated optical spectra (panels (b), (d), (f)). The red markers represent the estimated ERF.}
    \label{L_scan_alpha_maggiore1}
\end{figure}



\section{Transition between two different harmonic states}
\begin{figure}[h!]
\begin{center}
    \includegraphics[width=1\textwidth]{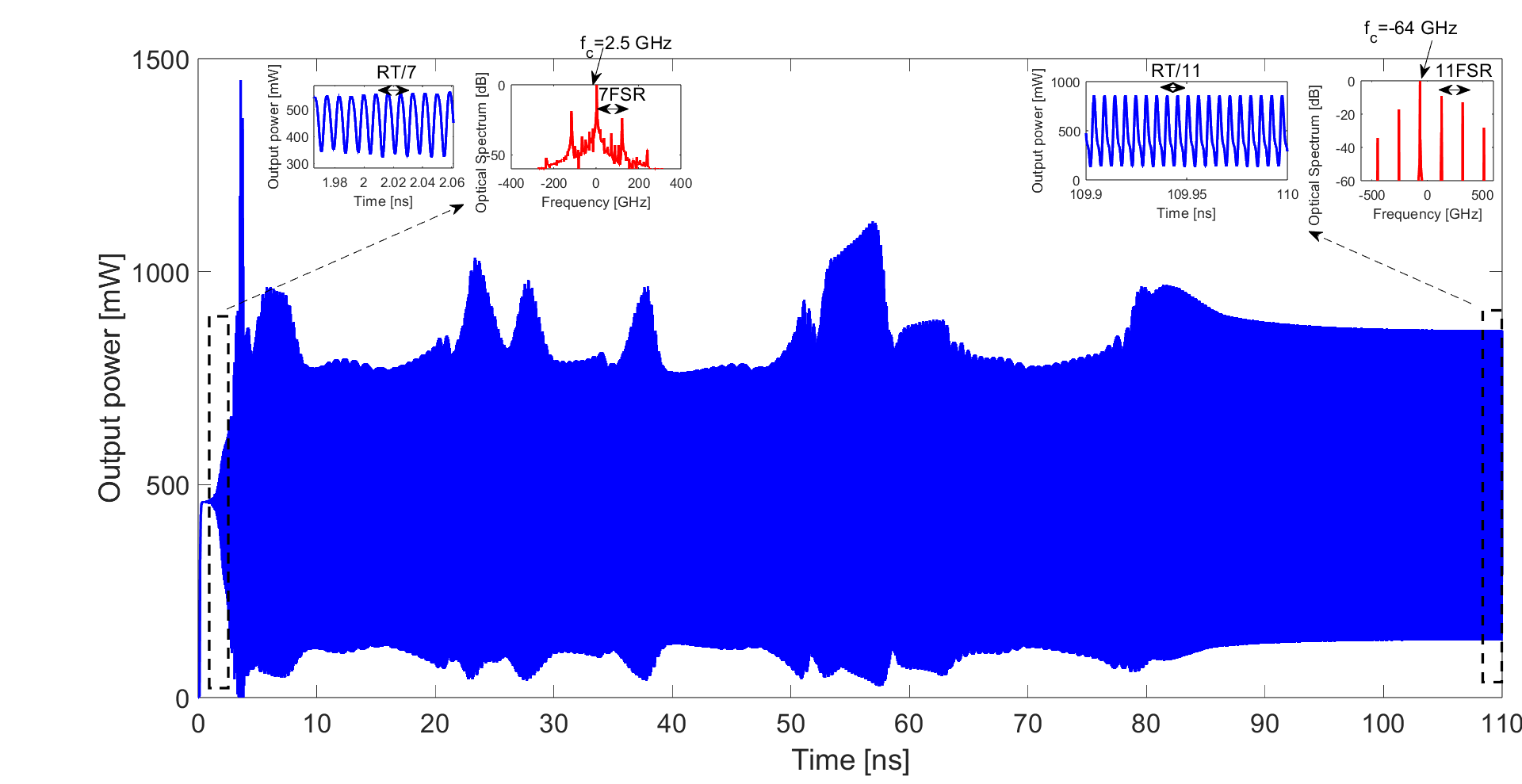}
       \end{center}

    \caption{Numerical simulation initially showing the formation of a transient 7th order HFC (insets on top left), followed by an irregular transient and the formation of a regular steady state 11th order HFC (insets on top right). Parameters: $\alpha=1.05$, $\Gamma=0.06$, $b=0.014$, and $\mu=5.9$.}
    \label{transition_fig}
\end{figure}
\begin{figure}
\begin{center}
    \includegraphics[width=0.5\textwidth]{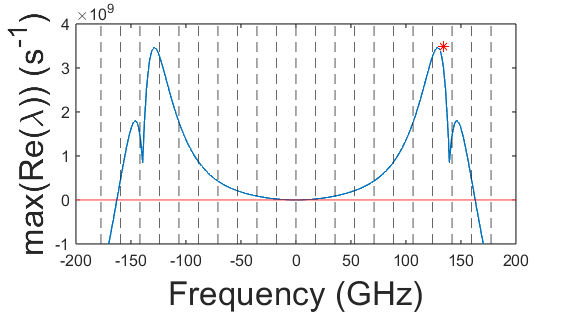}
       \end{center}

    \caption{Parametric gain and ERF (red marker) for the fundamental CW ($k=0)$ with the parameters of Fig.~\ref{transition_fig}. }
    \label{transition_fig_PG_ERF}
\end{figure}

We present here a simulation example in which a transition between two different harmonic states is observed for the same value of the pump parameter (see Fig.~\ref{transition_fig}). At the beginning of the simulation, the laser starts emitting in a continuous wave (CW) at $f_c = 2.5$ GHz from the reference frequency, which is the fundamental CW corresponding to $k=0$. This CW then destabilizes in favor of the formation of a 7th-order harmonic frequency comb (HFC) (see insets in the top left). We verified that the peak of the parametric gain and the ERF are in proximity to the cavity mode located 7 FSRs away from the reference frequency, as shown in Fig.~\ref{transition_fig_PG_ERF}.
After a few nanoseconds of simulation (Fig.~\ref{transition_fig}), a transition to an irregular transient occurs, lasting several tens of nanoseconds. Then, the system settles into a new 11th-order harmonic comb regime, with a central frequency of $f_c=-64$ GHz. Therefore, this second HFC corresponds to the destabilization of a different CW mode than the one associated with the harmonic comb observed in the first part of the simulation.


\section{Homoclons}
\begin{figure}[h!]
\begin{center}
    \includegraphics[width=0.75\textwidth]{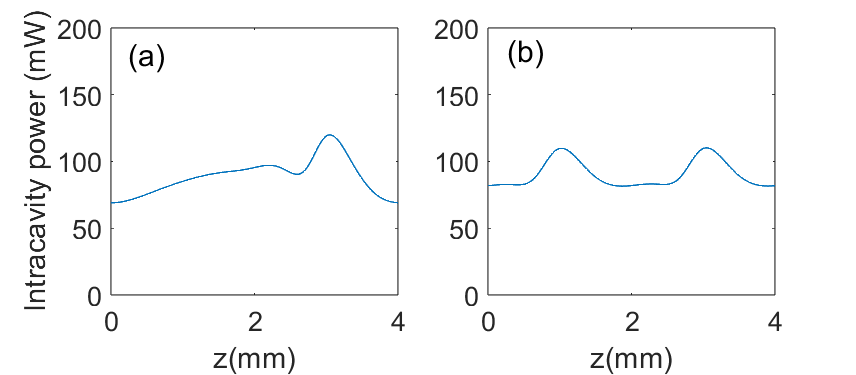}
       \end{center}
    \caption{Intracavity intensity profile at a fixed time, numerically obtained starting from two different initial conditions showing the formation of a single homoclon (panel (a)) or two homoclons (panel (b)) for $\alpha=1.5$, $\Gamma=0.1$, $b=0.02$, $\mu=2.15$ and $L=4$ mm.}
    \label{homoclon_fig}
\end{figure}

Homoclons are characteristic solitonic structures appearing in the complex Ginzburg-Landau equation (CGLE) in the parametric region where the Benjamin-Feir instability of the CW solution gives rise to phase-mediated turbulence \cite{Aranson} and they were experimentally observed in the presently studied class of devices in ref.~\cite{NaturePiccardo}. We verified their solitonic character in our simulations where we predict a Benjamin-Feir type instability ($\alpha > 1$). Close to the laser threshold, here two different initial conditions lead the system to reach steady states where one or two structures are present as plotted in Fig.~\ref{homoclon_fig}. The case of a single homoclon traveling in the cavity is associated with a dense OFC in the Fourier spectrum (line separation equal to a FSR).
This result is analogous to what was reported in \cite{Columbo2018} (see Fig.~8). 
\section{Damping coefficient in quantum well and quantum cascade lasers}
\begin{figure}[h!]
\begin{center}
    \includegraphics[width=0.4\textwidth]{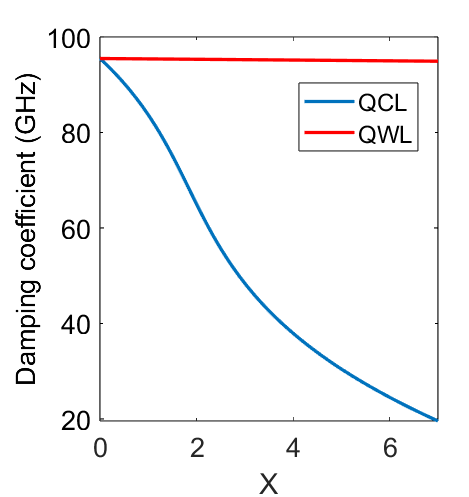}
       \end{center}
    \caption{Damping coefficient as a function of $X$ for a quantum cascade laser (QCL, blue) and a quantum well laser (QWL, red). The QCL parameters are $\alpha = 1.05$ and $b = 10^{-2}$, while the QWL parameters are $\alpha = 3$ and $b = 10^{-4}$. In both cases, $\Gamma = 0.06$.}
    \label{damping_fig}
\end{figure}

Figure~\ref{damping_fig} shows the damping coefficient, calculated as $|{\rm Re}(\beta)|/(2\pi\tau_d)$, as a function of $X$, obtained by solving the cubic characteristic equation~\eqref{beta_caratt} for the parameters of a QCL (blue curve) and a QWL (red curve). In the QCL case, the damping coefficient exhibits a pronounced decrease with increasing $X$. In contrast, for the QWL, it remains essentially constant and significantly higher than in the QCL, thereby inhibiting the observation of Rabi oscillations and harmonic frequency combs in QWLs.
\bibliography{supp}